\documentclass{article}
\usepackage{jheppub}
\usepackage{physics}
\usepackage{tikz}
\usepackage{comment}
\usepackage{upgreek}
\usepackage{makecell}
\usepackage{subcaption}

\newcommand\titlemath[1]{\texorpdfstring{#1}{Lg}}

\usepackage{xcolor}

\begin{document}

\title{
String stars, small black holes, and the supersymmetric index
}

\author[a,b]{Erez Y.~Urbach}
\affiliation[a]{School of Natural Sciences, Institute for Advanced Study, Princeton, NJ 08540, USA}
\affiliation[b]{Institute for Theoretical Physics, University of Amsterdam, Amsterdam, 1098XH, The Netherlands}

\emailAdd{urbach@ias.edu}

\abstract{
We discuss aspects of the black hole-string transition as the angular velocity approaches the supersymmetric index $\beta \Omega = 2\pi i$.
In this limit, the high-temperature Atick-Witten effective description breaks down.
For Heterotic string theory, a new high-temperature description emerges near the index in terms of a Higgsed $SU(2)_L\times U(1)_R$ supergravity with a magnetic field profile. The Hagedorn instability manifests in the theory as Ambjørn–Olesen W boson condensation, due to the background magnetic field.
For type II, a tower of winding strings becomes light near the index, which opens up a new dimension in the T-dual picture.
In both cases, a string star solution is found, similar to the Horowitz-Polchinski solution, but decouples at the index $\beta \Omega = 2\pi i$.\\
Considering charged solutions, the string star appears to interpolate between charged black holes and the free string phase. At the index, both the charged black hole and the charged string star phases decouple. This suggests a tension between a smooth black hole-string transition and the existence of a stringy BPS two-charge black hole.
}
\maketitle
\section{Introduction}
The high-energy spectrum of quantum gravity is dominated by black holes. In weakly coupled string theories, there is an intermediate high energy limit $1\ll l_s E \ll g_s^{-2}$ dominated by highly excited strings with a Hagedorn growth $S(E) \sim 2\pi R_H E$ \cite{Hagedorn:1965st,Huang:1970iq,Gross:1985fr,Bowick:1985az,Atick:1988si}. It is an old question \cite{Susskind:1993ws,Horowitz:1996nw,Peet:2000hn}, whether these two thermodynamic phases are secretly one and the same. Namely, if a high-temperature black hole (say in the last stages of evaporation) transitions to a self-gravitating string state, or a `string star'. This is sometimes referred to as the black hole/ string transition.
Since the two phases have no reliable overlapping region, attempts to argue for or against them have always been mostly qualitative.

This work studies the fate of the black hole/ string transition at the (target space) $1/2$-BPS supersymmetric index of either Heterotic or type II strings on $R^d$.
This index can be reached continuously from the thermal partition function, by turning on an angular velocity $\Omega$ at $\beta \Omega = 2\pi i$. 
For a given set of charges $Q$, the exact index $S_\text{index}(Q)$ is known \cite{Dabholkar:1989jt,Dabholkar:1990yf} by a free string theory calculation.
One may ask for the contribution of either the black hole or the string star to the index, by following them to the value $\beta \Omega = 2\pi i$. Since the index is independent of temperature and string coupling, agreement or disagreement between the two phases and the microscopic answer could shed light on the nature and validity of the string-black hole transition itself.

Unlike the black hole, it is not clear at all how to deform the string star saddle away from the thermal case. 
To understand why, let us recall what the string star saddle really is. We denote $R=\beta/(2\pi)$ the radius of the thermal circle.
Viewing the canonical partition function as a Euclidean string amplitude, a mode with a single winding along the thermal circle turns tachyonic for $R<R_H$, accounting for the Hagedorn growth of string states.
Upon reducing on the thermal circle, one obtains the Atick-Witten effective field theory (EFT) \cite{Atick:1988si}, which includes all the standard massless spectrum, together with a ``thermal scalar'': a complex scalar with mass $m^2 \sim R-R_H$. The string star saddle, or the Horowitz-Polchinski solution \cite{Horowitz:1997jc,Chen:2021dsw}, is a normalizable solution for the thermal scalar coupled to Newtonian gravity.\footnote{
A recent body of work on the subject includes rotating solutions \cite{Ceplak:2023afb,Ceplak:2024dxm,Santos:2024ycg,Seitz:2025wpc}, generalization to higher dimensions ~\cite{Balthazar:2022szl,Balthazar:2022hno,Bedroya:2024igb,Chu:2025boe,Chu:2025kzl}, and other background geometries and topologies ~\cite{Urbach:2022xzw,Urbach:2023npi, Halder:2023nlp,Agia:2023skp,
Ishibashi:2025qwn,
Chu:2024ggi,Emparan:2024mbp,Chu:2025fko}.}
The Atick-Witten EFT is reliable as long as the thermal scalar mass is light, which happens slightly below the Hagedorn temperature $R-R_H \ll 1$.
Perturbatively in $\Omega$, one can employ the Atick-Witten EFT to find slowly rotating string star solutions \cite{Seitz:2025wpc}.
But as we turn on $\Omega \sim O(1)$, the potential for the thermal scalar becomes string-sized, leading to an unreliable EFT. Thus, it appears hopeless to reliably deform the string star saddle to the index.

Surprisingly, although the Atick-Witten EFT breaks down, we find a novel EFT close to the supersymmetric point $\beta\Omega = 2\pi i$. In section \ref{sec:het_s1}, we begin with Heterotic string theory. At the index, the thermal circle admits a (periodic) supersymmetric boundary condition for the fermions. 
At the self-dual radius $R=1$ (in string units), the effective $d$-dimensional theory is half-maximal $SU(2)_L\times U(1)_R$ gauged supergravity. Away from the self-dual point, the theory is higgsed back to the standard KK $U(1)_L\times U(1)_R$ gauge group. This gauge enhancement is a well-known property of Heterotic string theory \cite{Narain:1985jj,Narain:1986am}.\footnote{In 
\cite{Balthazar:2022szl} (see also \cite{Balthazar:2022hno,Chu:2025boe,Chu:2025kzl}) a similar $SU(2)$ was considered on the worldsheet. 
In our work, the $SU(2)$ worldsheet currents are not projected out, and also exist in the target space as a gauge group.}
For temperatures close to the self-dual radius $R-1 \ll 1$, the EFT is reliable. The winding mode takes place in the EFT as the W boson. Due to supersymmetry, the W boson is stable, which agrees with the absence of a Hagedorn instability due to supersymmetric cancellations.
By going slightly away from $\beta \Omega = 2\pi i$, the Hagedorn instability is restored, and the theory should have a tachyon.
In the $SU(2)_L\times U(1)_R$ EFT, this instability has an elegant explanation.
Changing $\Omega$ amounts to a 
magnetic field profile for the unbroken $U(1)_L \times U(1)_R$ in the direction of the rotation plane. Since the W boson is charged, its mass is affected by the magnetic field. The result is known as the Ambjørn--Olesen phenomenon \cite{Ambjorn:1988tm,Ambjorn:1988fx,Ambjorn:1989bd}, in which the electroweak W boson condenses if the background magnetic field is high enough compared to the Higgs VEV. In our context, it explains how the winding mode (W boson) turns tachyonic away from the index (magnetic field) at high temperatures (small Higgs VEV).

In section \ref{sec:w_stars}, we look for string star solutions close to the index. In terms of the EFT, we work close to the critical magnetic field and look for a bubble of W boson condensation. We call those solutions `W stars'. Similar to the Horowitz-Polchinski solution, these solutions include a winding-mode profile as well as the gravitational (and magnetic) forces needed to support it. The solution shares many properties with the Horowitz-Polchinski solution, with a notable difference: the W star decouples as we tune back to the index $\beta \Omega = 2\pi i$. As we explain, this property is also shared by rotating black holes in this limit.

Next, in section \ref{sec:charged_saddles}, we use the method put forward in \cite{Chen:2021dsw} and consider charged solutions. Slightly away from the index, we find such charged W stars and study their properties. Similar to \cite{Chen:2021dsw}, our solutions appear to interpolate between the charged rotating black hole and the free string phase and approach the correct microscopic index $S_\text{index}(Q)$. This is drastically different than the charged black hole saddle, where a naive continuation to the index gives a vanishing index.
Exactly at the index, however, both the weakly curved charged black hole and the charged W star are unreliable and decouple from the spectrum.
Thus, we argue that a smooth black hole/string transition contradicts the existence of a BPS `small black hole' solution of the type recently suggested in \cite{Chowdhury:2024ngg,Chen:2024gmc}.

In section \ref{sec:type_ii} we comment on the type II case. Close to the index, an arbitrary number of winding strings become light at the same time. Thus, the EFT in this case is the T-dual $D=d+1$ dimensional SUGRA \cite{David:2001vm}. We argue that string star solutions exist with properties similar to the Heterotic W star. In section \ref{sec:bosonic} we consider Bosonic string theory, and explain a possible issue with the black hole/string transition when turning on large imaginary $\beta \Omega$. We close with conclusions and future directions.

\section{Heterotic compactifications near the self-dual radius}\label{sec:het_s1}
\subsection{Setup}\label{subsec:setup}
This work centers around the flat space string theory partition function
\begin{equation}\label{eq:Z_intro_def}
    Z(R,\nu) 
    = \text{Tr} 
    \left(
    e^{-2\pi R \hat H  - 2\pi i \nu J }(-1)^F 
    \right),
\end{equation}
with inverse temperature is $\beta = 2\pi R$, $H$ is the Hamiltonian, and $J$ is an angular momentum generator in the $x^1,x^2$ directions. 
At $\nu=0$, the partition function is the supersymmetric index, while $\nu$ introduces an imaginary angular velocity. Since $(-1)^F = \exp(2\pi i J)$, at $\nu=1$ \eqref{eq:Z_intro_def} coincides with the thermal partition function.\footnote{In \cite{Seitz:2025wpc} the definition of $\nu$ was such that $\nu=0$ is the thermal case. That is, $\nu_\text{there} = 1-\nu_\text{here}$.}

For now, we take the spatial geometry to be $R^9$, although later we will compactify it to $R^{d}$.
In \eqref{eq:Z_intro_def} the angular momentum generator $J$ is the rotation in the $x^1,x^2$ plane. 
The twisted partition function \eqref{eq:Z_intro_def} is defined by considering the partition function on top of the flat metric ($\rho^2 = (x^1)^2+(x^2)^2$, $X = x^3,...,x^9$)
\begin{equation}\label{eq:flat_metric}
\begin{split}
    ds^2 &= R^2 (dx^0)^2 + \rho^2 (d\theta')^2+d\rho^2+ dX^2,
\end{split}
\end{equation}
with the twisted boundary conditions $(x^0,\theta') \sim (x^0+2\pi, \theta' + 2\pi \nu)\sim (x^0,\theta'+2\pi)$ and periodic boundary condition for the fermions.\footnote{More precisely, in field theory, the trace \eqref{eq:Z_intro_def} is given by the partition function on top of the background \eqref{eq:flat_metric}. In string theory, it is defined by being asymptotically \eqref{eq:flat_metric}. Strictly speaking, these boundary conditions (besides at $\nu=0$) are not well-put in flat space and finite $g_s$ \cite{Atick:1988si}. One resolution would be to embed the discussion in anti-de Sitter space \cite{Urbach:2022xzw,Urbach:2023npi} or by considering (as we do in this section) the direct $g_s=0$ limit.} 

It will be useful to write the background in Kaluza-Klein form ($i,j=1,..,9$) 
\begin{equation}\label{eq:kk_red}
    ds^2 = g_{ij} dx^i dx^j+ e^\sigma \left(dx^0 + a_i dx^i\right)^2, B = B_{ij} dx^i \wedge dx^j + b_i dx^0 \wedge dx^i,
\end{equation}
with $B$ the Kalb-Ramond field. We further change the basis for the KK vectors to a ``left and right'' linear combinations
\begin{equation}\label{eq:A_bar_A}
    A_i = a_i+b_i, \quad 
    \bar A_i = a_i-b_i.
\end{equation}
By defining $\theta = \theta'-\nu x^0$, the background can be written as
\begin{equation}\label{eq:melvin_back}
\begin{split}
    g & = d\rho^2 + \frac{\rho^2}{1+\nu^2 \rho^2/R^2} d\theta^2 + dX^2,\\
    \sigma &= \log \left(R^2 + \nu^2 \rho^2\right), \\
    A = \bar A &= \frac{\nu \rho^2}{R^2+\nu^2 \rho^2} d\theta, \\
    \phi &= -\frac{1}{4} \log \left(R^2 + \nu^2 \rho^2\right),
\end{split}
\end{equation}
where we defined the $9$-dimensional dilaton $\phi = \Phi-\frac{1}{4}\sigma$. As we can see from a $9$-dimensional perspective, the background includes a non-trivial metric, dilaton, and graviphoton profile.
The latter should be thought of as a magnetic field $dA_{1,2} \ne 0$ localized around $\rho=0$. This is the Euclidean version of the magnetic Melvin background \cite{Melvin:1963qx,Gibbons:1987ps,Dowker:1993bt,Tseytlin:1994ei,Dowker:1995sg,Russo:1995tj,Gutperle:2001mb, David:2001vm}.

\subsection{Worldsheet analysis}\label{subsec:worldsheet}
\begin{figure}[ht]
    \centering
    \includegraphics[width=0.7\linewidth]{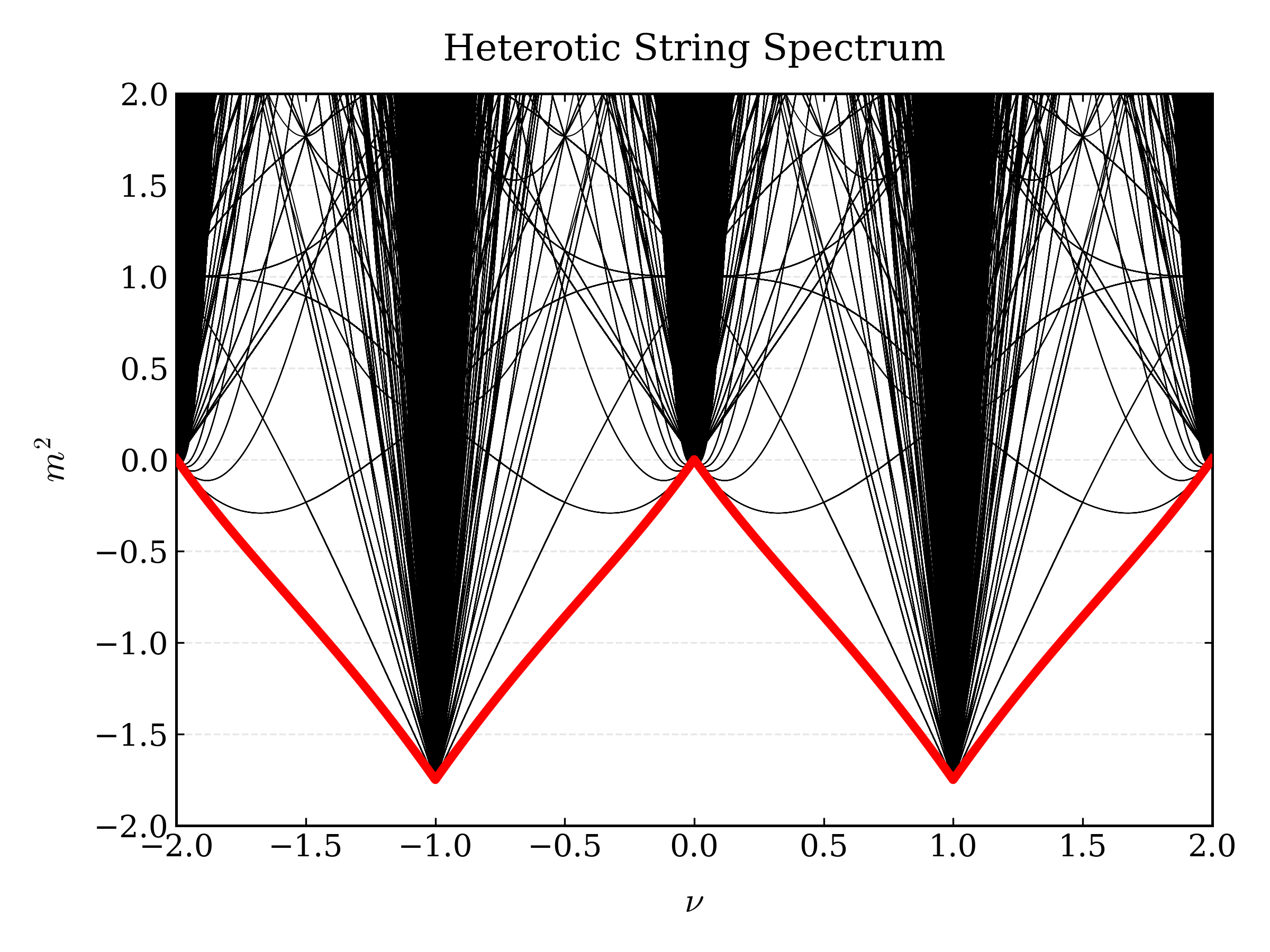}
    \caption{The Heterotic $(d-2)$-dimensional winding $w=1$ spectrum as a function of $\nu \sim \nu+2$. We set $R=1$ at which the classical contribution is zero. The spectrum is tachyonic (for small enough $R-1$) for any $\nu \ne 0 \text{ mod } 2$.
    The spectrum is dense close to $\nu=0$ and $\nu=1$, while for any $\nu$ the ground state (in red) is unique.}
    \label{fig:heterotic_spectrum}
\end{figure}

Unless said otherwise, we will use $\alpha'=1$ units throughout. 
The worldsheet theory that corresponds to \eqref{eq:Z_intro_def} is simply an orbifold of the flat space (free theory) worldsheet. Defining the complex worldsheet fields $Z = X^1 + i X^2 = \rho e^{i \theta'}$ and $\Psi = \psi^1 + i \psi^2$ ($\psi^\mu$ are the worldsheet fermions), the flat metric \eqref{eq:flat_metric} takes the form
\begin{equation}\label{eq:st_back}
    ds^2 = R^2 (dX^0)^2 + |dZ|^2 + \sum_{i=3}^{9} (dX^i)^2.
\end{equation}
In this form, we orbifold by the $\mathbb{Z}$ group action
\begin{equation}\label{eq:group_action}
    X^0 \mapsto X^0 + 2\pi, \quad Z \mapsto e^{2\pi i \nu} Z, \quad \Psi \mapsto e^{2\pi i \nu} \Psi.
\end{equation}

The spectrum of the orbifold theory decomposes into twisted/winding sectors labeled by the winding number $w$. 
Here we specialize to Heterotic string theory. 
For every $R$ and $\nu$ we would like to find the light spectrum of this background.
The low energies $w=0$ sector includes the standard SUGRA light spectrum of $10$d Heterotic strings around the background \eqref{eq:melvin_back}.
By tuning $R$, the first winding sector $w=\pm1$ can also be light. 
To see that, we write the mass-shell conditions for $w=1$  and $0\le \nu\le 1$ in the NS sector\footnote{The R-sector ground state for $w>0$ can be shown always to be more massive than the NS ground state.} \cite{Seitz:2025wpc}
\begin{equation}\label{eq:mass_shell}
\begin{split}
    m^2 &= (R+n/R)^2-4 + 2 \nu (1-\nu) + 4 N \\
    &=(R-n/R)^2-1-|2\nu-1| + 4 \tilde N,
\end{split}
\end{equation}
with $n$ being the momentum along the thermal circle, and $N,\tilde N$ the left/right-moving levels. For brevity, we choose $w=1$ although the same spectrum (with opposite $n$) exists also for $w=-1$.
What is the lowest allowed mass in this sector? Since the GSO projection doesn't touch the left movers, we can always set $N=0$.
In the thermal case $\nu=1$, the GSO projection allows for $\tilde N=0$ as well, which gives the mass and thermal momentum \cite{Atick:1988si}
\begin{equation}\label{eq:het_mass}
    m^2_{\nu=1} = R^2 + \frac{1}{4R^2}-3, \quad n_{\nu=1} = \frac{1}{2}.
\end{equation}
A simple calculation shows that the mass turns tachyonic for $R<R_H = \frac{1}{2}(2+\sqrt{2})$, the Heterotic Hagedorn temperature \cite{Gross:1985fr}. By tuning the temperature slightly above the Hagedorn temperature $R-R_H \ll 1$, the winding mode can be made arbitrarily light (in string units) $m^2 \ll 1$. In the ``near-Hagedorn'' temperature regime $R-R_H \ll 1$, the $9$-dimensional EFT thus includes both the $9$-dimensional KK massless Heterotic spectrum and a single complex scalar field with mass \eqref{eq:het_mass}. This is the Atick-Witten EFT \cite{Atick:1988si}.

Away from $\nu=1$, the rotation-plane fields  $Z,\Psi$ can no longer have a center-of-mass momentum in the rotation plane. Instead, the twisted boundary conditions (for $w=1$)
\begin{equation}\label{eq:twisted_BC}
    Z(\sigma + l) = e^{2\pi i \nu} Z(\sigma), \quad \Psi(\sigma + l) = -e^{2\pi i \nu} \Psi(\sigma),
\end{equation}
break translation-invariance and
shifts the standard harmonics of $Z$ ($\Psi$) from $\omega_n = |n|$ ($|n-1/2|$) to $\omega_n = |n+1-\nu|$ ($|n+1/2-\nu|$). The lowest allowed levels are (integer products of) $\tilde N \sim |1-\nu|$ (see appendix \ref{app:thermal_eft} for a target space derivation in this regime).
The characteristic energy gap is therefore $\Delta m^2 \sim |1-\nu|$ (in string units).
Thus, the Atick-Witten EFT remains meaningful only for $|1-\nu| \ll 1$.  See figure \ref{fig:heterotic_spectrum}.

Although the Atick-Witten EFT is no longer valid, at the supersymmetric index $\nu=0$ $Z$, $\Psi$ are no longer twisted \eqref{eq:twisted_BC}, the spectrum is dense again, and one can hope for a new $9$-dimensional EFT.
If we follow the thermal winding mode as we change $\nu$, the right-moving level flows as \cite{Seitz:2025wpc}
\begin{equation}
    \tilde N = \begin{cases}
        \frac{1}{2}-\nu, & 0\le \nu\le \frac{1}{2}\\
        0 & \frac{1}{2} < \nu \le 1
    \end{cases}.
\end{equation}
The discontinuity in $\tilde N$ is due to the fact that the ground state changes its fermionic number, which changes the GSO projection.
For $0<\nu<\frac{1}{2}$, the level comes from the fractional $\Psi$ oscilator with frequency $\omega = |1/2-\nu|$.
Right at the supersymmetric point $\nu=0$ the right-moving level is $\tilde N = \frac{1}{2}$, corresponding via \eqref{eq:mass_shell} to a mass and thermal momentum
\begin{equation}\label{eq:susy_vector_mass}
    m^2_{\nu=0} = (R-1/R)^2, \quad n_{\nu=0} = 1.
\end{equation}
The level $\tilde N = 1/2$ is the standard NS excitation, with the vertex operator $V^\pm_0 = e^{\pm 2i X_L} \psi^{0}$. Since the same level and mass can be achieved by using any of the other worldsheet fermions $V^\pm_i = e^{\pm 2i X_L} \psi^{i}$ ($i=1,...,8$), this is not a target-space scalar anymore, but combines into a massive (complex) vector field!

What is going on? Contrary to the thermal case \eqref{eq:het_mass}, the vector's mass \eqref{eq:susy_vector_mass} is non-negative for any $R$. This should not surprise us, since we don't expect any tachyonic instabilities or Hagedorn growth of states for the supersymmetric index. It does, however, turn massless at $R=1$, where Heterotic string theory has an enhanced gauge symmetry, from the $U(1)_L \times U(1)_R$ momentum and winding to $SU(2)_L \times U(1)_R$ \cite{Narain:1985jj,Narain:1986am}.\footnote{See \cite{Aldazabal:2017jhp,Fraiman:2018ebo} for a recent discussion of this subject and references therein.}
At $R=1$, the vertex operators $V_0^\pm$ and $V_i^\pm$ correspond to a $9$-dimensional massless vector multiplet made of a scalar and vector, respectively. The two new gauge vectors correspond to the enhancement of the left-moving $U(1)$ to $SU(2)$ at $R=1$.
In terms of \eqref{eq:kk_red}, the massless Bosonic spectrum at $R=1$ include ($\alpha=1,...,3$)
\begin{itemize}
    \item Vectors in the algebra of $SO(32)$ or $E_8\times E_8$, and their KK reduction. Since we set all thermal holonomies to zero, this sector will decouple from our discussion and will not concern us further.
    \item $9$-dimensional gravitational sector: the graviton $g$, antisymmetric tensor $B$, and dilaton $\Phi$.
    \item A right-moving KK abelian gauge field $\bar A_i$ \eqref{eq:A_bar_A}, the generator of $U(1)_R$.
    \item An $SU(2)_L$ gauge field $A^\alpha_i$, composed of the left-moving KK vector $A^3_i = A_i$ \eqref{eq:A_bar_A}, and two new fields $A^\pm_i$ with vertex $V^\pm_i$.
    \item An $SU(2)_L$ adjoint Higgs scalar $S^\alpha$. It is composed of the KK-scalar $S^3 =\sigma$, and two $S^{\pm}$ with vertex $V^\pm_0$.
\end{itemize}
The adjoint Higgs has no potential and allows for any VEV. At the self-dual radius, the Higgs has no VEV, and the massless spectrum includes the $SU(2)_L\times U(1)_R$ gauge fields and scalars. 
The Higgsed sector of the theory corresponds to changing the radius away from $R=1$. 
By the Higgs mechanism, two (spontaneously broken) $SU(2)_L$ gauge fields combine with two scalars into a complex massive W vector boson. This is exactly the massive vector that corresponds to $V^\pm_0$ and $V^\pm_i$. Thus, the Atick-Witten winding mode is identified in the $SU(2)_L\times U(1)_R$ theory as the W boson.

Finally, let us show that the $SU(2)_L\times U(1)_R$ theory should remain consistent also for small angular velocity $\nu \ll 1$. 
The continuous $x^1,x^2$ momenta are discretized at finite $\nu \ne 0$ by the twisted boundary conditions \eqref{eq:twisted_BC}. For small $\nu$, the light spectrum comes from levels of the type ($k,\tilde k \in \mathbb{Z}$)
\begin{equation}
    N = k \ \nu, \quad \tilde N = \frac{1}{2} + \nu\  \tilde k.
\end{equation}
As we discussed above, the ground state is $k=0$, $\tilde k = -1$, and otherwise $k\ge 0$, $\tilde k\ge -1$. Considering only ``transverse modes'' (with $\psi^i$ $i\ne 1,2$), the lowest right-moving level is $\tilde N = \frac{1}{2}$, and therefore $\tilde k \ge 0$. Solving \eqref{eq:mass_shell}, gives the mass and thermal momenta
\begin{equation}\label{eq:m2_n_nt}
    \begin{split}
        m^2 & = R^2 + \frac{1}{R^2}\left(1+\nu (\tilde k -k)+\nu^2/2\right)^2 -2+2(1+k+\tilde k)\nu-\nu^2
        , \quad n = 1+(\tilde k-k) \nu + \nu^2/2\\ 
        &= (R-1/R)^2 + 2\nu\left(1 + k+\tilde k + \frac{\tilde k-k}{R^2}\right) + O(\nu^2),
    \end{split}
\end{equation}
with a characteristic energy difference $\Delta m^2 \sim \nu$. Therefore, scale separation of the EFT remains as long as $\nu \ll 1$. The winding ground state $k=0$, $\tilde k = -1$ satisfies
\begin{equation}\label{eq:het_m2}
    m^2 = (R-1/R)^2 - \frac{2 \nu}{R^2} + O(\nu^2), \quad n = 1-\nu + O(\nu^2).
\end{equation}
In the target space, this is the effective mass of the W boson in the longitudinal directions $x^1,x^2$.
At $\nu=0$ we get \eqref{eq:susy_vector_mass} again. However, we see that turning on $\nu > 0$ reduces the mass, allowing a tachyonic instability to arise. The ground state is tachyonic for $R<R_H(\nu)$ with 
\begin{equation}
    \begin{split}
        R_H^2(\nu) &= 1+\sqrt{\nu(2+\nu(\nu-1))}+\nu^2/2\\
        &= 1+\sqrt{2\nu} + O(\nu).
    \end{split}
\end{equation}
\newline

Let us recap. We followed the thermal-scalar winding mode from the thermal case to the index $\nu=0$. 
Although the original Atick-Witten EFT was valid only for $|1-\nu|\ll 1$, close to the index $\nu \ll1$, a new half-maximal $SU(2)_L\times U(1)$ SUGRA description exists for $R-1 \ll 1$. 
In the new EFT, the light winding modes are organized in terms of a massive W vector boson. Put differently, `W' stands here for winding!\footnote{
In general, there's no relation between the winding scalar field $\chi(x)$ in the Atick-Witten theory and the W boson in the $SU(2)$ theory. Some stringy modes turn light at $\nu=1$ and others at $\nu=0$, and one should not be tempted to relate the two EFTs directly. However, as we showed, the winding-sector ground state is the same in both theories. See figure \ref{fig:heterotic_spectrum}.}
In that theory, the conditions $R-1 \ll 1$ and $\nu \ll 1$ are equivalent to a small VEV for the Higgs and a small magnetic field profile for the unbroken $U(1)_L\times U(1)_R$. In the following sections, we will write down this theory and study it.

\subsection{The \titlemath{$SU(2)$} SUGRA theory}\label{subsec:su2_sugra}
Let us begin with the Heterotic 10d low-energy effective action in Euclidean coordinates
\begin{equation}
    I_E = \frac{1}{16\pi G_N} \int d^{10} x e^{-2\Phi} \sqrt{G} \left(-\mathcal{R}^{(10)} 
    - 4 (\partial \Phi)^2 + \frac{1}{12}H^2
    \right),
\end{equation}
with $H^2 = H_{ijk} H^{ijk}$. We didn't write the $SO(32)$ (or $E_8 \times E_8$) sector since we consistently turn it off in our analysis.
Assuming an $R^9\times S^1_R$ ($x^0 \sim x^0 +2 \pi$) topology, the general KK reduction \eqref{eq:kk_red}\footnote{We use a different convention than \cite{Polchinski:1998rq}, with $\sigma = 2\sigma_\text{Polchiski}$.}
together with $\phi = \Phi-\frac{1}{4}\sigma$ gives the 9d effective action
\begin{equation}\label{eq:KK_1}
    I_E = \frac{1}{8 G_N} \int d^{9} x e^{-2\phi} \sqrt{g} \left(-\mathcal{R}^{(9)} +\frac{1}{12}H^2
    - 4 (\partial \phi)^2 + \frac{1}{4}(\partial \sigma)^2 
    + \frac{1}{4} e^{\sigma} f^2 + \frac{1}{4} e^{-\sigma} h^2
    \right),
\end{equation}
with $f=da$, $h=db$.
For the next step, we move to ``left and right'' vector field basis \eqref{eq:A_bar_A}.
In terms of $A_i$, $\bar A_i$, the action becomes ($F=dA$, $\bar F=d\bar A$) \footnote{We have a different convention for the root lattice compared to \cite{Fraiman:2018ebo}, hence the different overall $1/g_{YM}^2$.}
\begin{equation}\label{eq:KK_2}
\begin{split}
    I_E &= \frac{1}{8 G_N} \int d^{9} x e^{-2\phi} \sqrt{g} \left(-\mathcal{R}^{(9)} +\frac{1}{12}H^2
    - 4 (\partial \phi)^2 \right.\\
    &\left. + \frac{1}{4}(\partial \sigma)^2 
    + \frac{1}{8} \cosh(\sigma) \left(F^2+\bar F^2\right) + \frac{1}{4} \sinh(\sigma) F_{ij} \bar F^{ij}
    \right).
\end{split}
\end{equation}

The circle reduction preserves all 16 supercharges of the Heterotic strings. Thus, the $D=9$ theory we are looking for is a half-maximal $\mathcal{N}=1$ SUGRA.
In terms of $\mathcal{N}=1$ multiplets, the gravity multiplets include both $g$, $B$, $\bar A$ and $\phi$. The left-moving vector $A$ combines with $\sigma$ to form their own vector multiplets.
Recall that the winding mode has $w=n=\pm1$ \eqref{eq:susy_vector_mass}, and therefore charged under the left-moving $A_i$ and neutral under $\bar A_i$. Thus, we need to introduce two new vector multiplets made of two gauge and scalar fields $A^{1,2}$, $S^{1,2}$, which will together combine with $A_i,\sigma$ into an $SU(2)$ vector multiplet.
As we explain in appendix \ref{app:sugra}, such $9$d $\mathcal{N}=1$ SUGRA is fixed to have the following Bosonic action up to two derivatives and all orders in the fields:
\begin{equation}\label{eq:su2_cov}
\begin{split}
    I_{SU(2)} &= \frac{1}{8 G_N} \int d^{9} x e^{-2\phi} \sqrt{g} \left(-\mathcal{R}^{(9)}  +\frac{1}{12}H^2
    - 4 (\partial \phi)^2 \right.\\
    &+
    \frac{1}{4} (D S)^2 + \frac{1}{4}\left(\frac{2\cosh|S|-2}{S^2} -1\right) [D S,\hat S]^2\\
    &\left.+ \frac{1}{8}\cosh|S|\left( F^2+ \bar F^2\right) + \frac{1}{8} \left(1-\cosh|S|\right) [F,\hat S]^2 + \frac{1}{4} \sinh|S| (\hat S \cdot F) \bar F
    \right),
\end{split}
\end{equation}
Here $A^\alpha$, $S^\alpha$ ($\alpha=1,2,3$) are adjoint $SU(2)$ gauge field and scalar, respectively, in a basis $T^\alpha = \frac{1}{2}\sigma^\alpha$ with structure constants $f^\alpha_{\beta \gamma} = \epsilon^\alpha_{\beta \gamma}$. We denoted $\hat S^\alpha = S^\alpha/|S|$, $|S| = \sqrt{S^2}$ and 
\begin{equation}
    F^\alpha_{ij} = \partial_i A^\alpha_j - \partial_j A^\alpha_i + \epsilon^\alpha_{\beta \gamma} A^\beta_i A^\gamma_j, \quad
    D_i S^\alpha = \partial_i S^\alpha + \epsilon^\alpha_{\beta \gamma} A^\beta_i S^\gamma.
\end{equation}
All the spacetime and color index summations in \eqref{eq:su2_cov} are implicit, for example $(\hat S \cdot F) \bar F = \hat S^\alpha F^\alpha_{ij} \bar F^{ij}$.
Note that this theory contains no potential for the scalar $S^\alpha$.

Throughout the analysis, we will use the unitary gauge, in which we set two directions of $S^\alpha$ to zero
\begin{equation}\label{eq:unitary_gauge}
    S = \sigma \ T^3.
\end{equation}
To make contact with the KK theory, it is also convenient to rename 
\begin{equation}\label{eq:W_def}
    A \equiv A^3, \quad 
    W \equiv \frac{1}{\sqrt{2}} (A^1+i A^2),\quad
    W^* \equiv \frac{1}{\sqrt{2}} (A^1-i A^2).
\end{equation}
Due to the $SU(2)$ structure, $W$ is a vector with charge $+1$ under $U(1)_L$, generated by the gauge field $A_i$.
In this gauge, the theory takes the simpler form
\begin{equation}\label{eq:W_su2_action}
\begin{split}
    I_{SU(2)} &= \frac{1}{8 G_N} \int d^{9} x e^{-2\phi} \sqrt{g} \left(-\mathcal{R}^{(9)}  +\frac{1}{12}H^2
    - 4 (\partial \phi)^2 \right.\\
    &+
    \frac{1}{4} (\partial \sigma)^2 
    + \frac{\cosh(\sigma)}{8}\left(F^2+ \bar F^2\right)+ \frac{\sinh(\sigma)}{4} F_{ij} \bar F^{ij}
    \\
    &  + \frac{1}{4}|DW|^2+ \frac{2\cosh(\sigma)-2}{2} |W|^2\\
    &\left.
    + \frac{i}{4} \left(\cosh(\sigma) F^{ij}+\sinh(\sigma) \bar F^{ij}\right) W_{[i}W^*_{j]}
    +\frac{\cosh(\sigma)}{8} \left|[W,W^*]\right|^2
    \right),
\end{split}
\end{equation}
with $DW = dW + i A \wedge W$, $D W^* = d W^* - i A \wedge W^*$, $F = dA$ and $[W,W^*]_{ij}= W_{[i} W^*_{j]}$.
As we turn off $W=0$, we get back the KK theory \eqref{eq:KK_2}, so that \eqref{eq:W_su2_action} is indeed a gauge enhancement of the KK reduction. While this $SU(2)_L\times U(1)_R$ was written down before up to cubic order \cite{Aldazabal:2017jhp,Fraiman:2018ebo} (see also \cite{Aldazabal:2015yna}), as far as we know this is the first time \eqref{eq:W_su2_action} is written to all fields order up to two-derivatives (see appendix \ref{app:sugra}).

As a simple test, consider the untwisted $R^d \times S^1_R$ background, which in KK form is
\begin{equation}\label{eq:S1_back}
    g_{ij} = \delta_{ij},\quad \sigma = \log R^2, \quad \phi = -\frac{1}{2}\log R, \quad A_i=\bar A_i=0. 
\end{equation}
This is a flat background geometry together with (for $R\ne 1$) a VEV for the adjoint Higgs. Such a background is a solution since $\sigma$ has no potential (at $A_i=\bar A_i = 0$).\footnote{Since $S^\alpha$ is part of a vector multiplet, supersymmetry jargon refers to its modulus as the Coulomb branch, not the Higgs branch. Nevertheless, we will follow particle-physics convention and refer to $S^\alpha$ as the Higgs.}
What is the spectrum of the Higgsed theory? As we can immediately read from \eqref{eq:W_su2_action}, the fields $\sigma, A_i, \bar A_i$ remain quadratically massless. Due to the Higgs mechanism, the VEV gives a mass to the W boson.
Plugging the background \eqref{eq:S1_back} in the quadratic piece for $W$ \eqref{eq:W_su2_action}, renders a mass
\begin{equation}
    m^2_W = 2\cosh \sigma -2 = (R-1/R)^2,
\end{equation}
matching \eqref{eq:susy_vector_mass}. Despite the exact match, note that the EFT is only valid as long as its spectrum is light $R-1\ll 1$. By contrast to the Atick-Witten EFT, the $SU(2)$ EFT holds to any order in the fields, so we don't have to assume that their amplitudes are small.

\subsection{The magnetic spectrum}\label{subsec:mag_spectrum}
Since the index's background \eqref{eq:S1_back} is supersymmetric, its spectrum is non-negative for any temperature $R$. In terms of the partition function \eqref{eq:Z_intro_def}, it signals no Hagedorn growth for the index. 
At the same time, as we saw in \eqref{eq:het_m2}, the high-temperature winding spectrum is tachyonic for $\nu >0$.
In terms of the $SU(2)$ theory, turning on $\nu$ (at small enough $R$) should destabilize the W boson vacuum. The goal of this section is to explain this instability by studying the spectrum of the W boson on top of the magnetic Melvin background \eqref{eq:melvin_back},
\begin{equation}\label{eq:melvin_back_pm}
\begin{split}
    g & = dz d\bar z + \nu^2 \frac{R^2}{4\Lambda^2} (z d\bar z -\bar z dz)^2 + dX^2,\\
    \sigma &= \log \Lambda^2, \\
    A = \bar A &= \frac{i \nu}{2\Lambda^2} \left(z d\bar z - \bar z dz\right), \\
    \phi &= -\frac{1}{4} \log \Lambda^2,
\end{split}
\end{equation}
with the complex coordinates $z,\bar z = x^1 \pm i x^2$ and $\Lambda^2 = R^2 + \nu^2 |z|^2$.
To quadratic order in $W$, $SU(2)$ theory \eqref{eq:W_su2_action} has the following form (since we only consider eigenmodes of the quadratic action, we omitted the overall factor for brevity)
\begin{equation}\label{eq:pert_action_2}
\begin{split}
    I_W^{(2)} &= \int d^9 x \left(
    \frac{1}{4}|DW|^2 
    + \frac{2\cosh(\sigma)-2}{2} |W|^2
    + \frac{i}{4}\left(\cosh(\sigma) F + \sinh(\sigma) \bar F\right) [W, W^*]
    \right)\\
    &= \int d^9 x \left(
    \frac{1}{4}|dW+i A \wedge W|^2 
    + \frac{(\Lambda-1/\Lambda)^2}{2} |W|^2
    + \frac{i}{2}\Lambda^2 F^{ij} W_{i} W^*_{j}\right),
\end{split}
\end{equation}
where in the second line, we used that $A_j=\bar A_j$,
and that the background's volume $e^{-2\phi}\sqrt{g}=1$ is trivial (the metric contractions are still non-trivial). 
The third term in \eqref{eq:pert_action_2}, contrary to the first two, is not positive semidefinite.
As will become clearer shortly, this term is responsible for destabilizing the system.
\newline

As a warm-up, let us consider transverse polarizations of W, of the type
\begin{equation}\label{eq:trans_ansatz}
    W = e^{-i p_i x^i} w(z,\bar z)\ dx^j,
\end{equation}
with a fixed transverse direction $j\ne 1,2$. By the gauge constraint \eqref{eq:gauge_const}, we also have $p_j = 0$ (or $p \cdot W = 0$). Because the momentum and the polarization are orthogonal to the magnetic field profile, $w$ behaves like a scalar: $(DW)_{k,j} = (\partial_k + i A_k) W_j$. For the same reason, the Zeeman term in \eqref{eq:pert_action_2} vanishes, leaving the quadratic action
\begin{equation}
\begin{split}
    I_W^{(2)} 
    &= \int d^2 z \left(
    \frac{1}{2} |(\partial + i A) w|^2 + \frac{1}{2}p^2 |w|^2
    +\frac{1}{2}\left(\Lambda-1/\Lambda\right)^2 |w|^2\right).
\end{split}
\end{equation}
This is the action of a scalar coupled to a gauge field, albeit in a nontrivial background and a position-dependent mass term. Such a theory is akin to the thermal situation around $1-\nu \ll 1$, in which the winding mode is a scalar (see appendix \ref{app:thermal_eft} for a review). 

To go forward, our goal is to find the spectrum up to sub-leading order $O(\nu)$. 
As will be clearer below, $w(z,\bar z)$ is localized to $z=\bar z = 0$ with a characteristic two-dimensional length scale $|z|^2 \sim 1/\nu$. 
Expanding the action up to order $O(\nu^2)$, amounts to  expanding the background metric, gauge field $A$, and mass-term by
\begin{equation}\label{eq:melvin_back_approx}
\begin{split}
    g & = dz d\bar z + dX^2,\\
    A &= \frac{i \nu}{2R^2} \left(z d \bar z - \bar z dz\right), \quad F =  \frac{i \nu}{R^2} dz \wedge d\bar z,\\
    (\Lambda-1/\Lambda)^2 &= (R-1/R)^2 + \nu^2 \left(1-R^{-4}\right) |z|^2.
\end{split}
\end{equation}
After some algebra, the $O(\nu)$ action takes the following form
\begin{equation}\label{eq:trans_simple}
\begin{split}
    I_W^{(2)} 
    = \int d^2 z &\left(
    |(\partial_z + i \hat A_z) w|^2 + |(\partial_{\bar z} + i \hat A_{\bar z}) w|^2 + \frac{1}{2}p^2 |w|^2
     +\frac{1}{2}\left(R-1/R\right)^2 |w|^2 \right.\\
    & \  \left.- \frac{\nu}{2}(1-R^{-2})\left(w^*(z \partial_z - \bar z  \partial_{\bar z})w + c.c.\right)\right) + O(\nu^2),
\end{split}
\end{equation}
with the auxiliary gauge field 
\begin{equation}
    \hat A = i \frac{\nu}{2} (z d\bar z - \bar z dz), \quad \hat F = i \nu \ dz \wedge d\bar z.
\end{equation}
The first line of \eqref{eq:trans_simple} is simply a scalar in a constant magnetic field $\hat F$ in flat space. The eigenbasis for this system is simply the Landau levels.
Because the second line is proportional to the orbital angular momentum $J = w^*(z \partial_z - \bar z  \partial_{\bar z})w$, the same Landau levels are still an eigen basis for the system. Let us recall the Landau analysis. The two pairs of creation and annihilation operators
\begin{equation}\label{eq:nu_op}
\begin{split}
    \tilde a = \frac{\nu}{2} \bar z + \partial_z, &\quad 
    \tilde a^\dagger = \frac{\nu}{2} z-\partial_{\bar z}, \\
    a = \frac{\nu}{2} z + \partial_{\bar z}, & \quad 
    a^\dagger = \frac{\nu}{2} \bar z -\partial_z, 
\end{split} 
\end{equation}
satisfy the commutation relations
\begin{equation}\label{eq:comm_rel}
    [a, a^\dagger] = [\tilde a, \tilde a^\dagger] = \nu,
\end{equation}
while the rest of the commutators are trivial.
In terms of these operators, the quadratic action is the expectation value
\begin{equation}\label{eq:trans_I2}
    I_2 = \int d^2 z \frac{1}{2} w^* \hat H w,
\end{equation}
of the following `Hamiltonian'
\begin{equation}\label{eq:H_def}
\begin{split}
    \hat H 
    &= p^2 + (R-1/R)^2 + 2\tilde a^\dagger \tilde a + 2\tilde a \tilde a^\dagger - 2(1-R^{-2})\left(\tilde a^\dagger \tilde a- a^\dagger a\right)\\
    &= p^2 +\left(R-1/R\right)^2 + 2\nu 
    + 2(1-R^{-2}) a^\dagger a + 2(1+R^{-2})\tilde a^\dagger \tilde a.
\end{split}
\end{equation}
Define the ground state wave-function
\begin{equation}
    \label{eq:w_gs_wavefunction_nu}
    w_{0,0}(z,\bar z) = \sqrt{\frac{\nu}{\pi}} \exp(-\nu |z|^2/2),
\end{equation}
to annihilates both $a \ w_{0,0} = \tilde a \ w_{0,0} = 0$. The rest of the Landau levels are constructed using the creation operators
\begin{equation}
    w_{k,\tilde k} = \frac{(a^\dagger)^{k}(\tilde a^\dagger)^{\tilde k}}{\sqrt{k! \tilde k!}} w_{0,0},
\end{equation}
with $k,\tilde k\ge 0$ and together form a complete basis. The corresponding eigenvalues read
\begin{equation}\label{eq:lambda_trans}
    \lambda_{k,\tilde k} = p^2 + (R-1/R)^2 + 2\nu \left(1+(1-R^{-2}) k + (1+R^{-2})\tilde k\right),
\end{equation}
which reproduce the worldsheet computation \eqref{eq:m2_n_nt}.
The wavefunction \eqref{eq:w_gs_wavefunction_nu} indicates that a characteristic size $|z|^2 \sim \nu^{-1}$, which justified the expansion above \eqref{eq:melvin_back_approx}. By dimensional analysis, the next order in $\nu$ gets corrections from higher-derivative terms. Thus, using the two-derivative action alone, this is the most we can do with our two-derivative EFT.
\newline

Since $k,\tilde k\ge 0$, the resulting spectrum is always non-negative $\lambda_{k,\tilde k} \ge 0$. This is no surprise, since we saw in section \ref{subsec:worldsheet} that the tachyon is a longitudinal (or in-plane) mode $W_{1,2}$, to which we turn next. In the presence of a magnetic field (see appendix \ref{app:adjoint_higgs}), these modes couple at quadratic order to a transverse ``pure-gauge'' mode of $W$. The general ansatz under the symmetries of the problem is (all the sums are implicitly over $j \ne 1,2$)
\begin{equation}\label{eq:long_ansatz}
    W = e^{-i p\cdot x} \left(w_+ dz + w_- d\bar z - i p_j dx^j \cdot \phi \right),
\end{equation}
with $w_\pm$ and $\phi$ all functions of $z,\bar z$. 
Let us write the different terms of the action already in the expanded background \eqref{eq:melvin_back_approx}
\begin{align}
    (DW)_{z,\bar z} &= D_z w_- - D_{\bar z} w_+\\
    (DW)_{z,j} &= -i p_j \left(D_z \phi - w_+\right)\\
    (DW)_{\bar z,j} &= -i p_j \left(D_{\bar z} \phi - w_-\right),\\
    \frac{1}{2}(\Lambda-1/\Lambda)^2 |W|^2 
    &= \left((R-1/R)^2 + \nu^2 (1-R^{-4})|z|^2\right) \left(
    |w_+|^2 + |w_-|^2
    + \frac{1}{2} p^2 |\phi|^2
    \right),\\
    \frac{i}{2} e^\sigma F^{ij} W_{i} W^*_{j} &= 
    2\nu (|w_+|^2-|w_-|^2),
\end{align}
where on the RHS $D_j = \partial_j + i A_j$.
Together, the quadratic action for the ansatz takes the form (upon integration by parts)
\begin{equation}
\begin{split}
    I_W^{(2)} = \int d^2z &
    \left\{
    2 \left|D_z w_- - D_{\bar z} w_+\right|^2 
    \right. \\
    &
    + p^2\left(|D_z \phi|^2 + |D_{\bar z} \phi|^2 
    + \phi^* (D_z w_- + D_{\bar z} w_+) + c.c.
    + \frac{1}{2}(R-1/R)^2 |\phi|^2
    \right)
    \\
    & 
    + (p^2 + (R-1/R)^2 + 2\nu)|w_+|^2
    + (p^2 + (R-1/R)^2 - 2\nu)|w_-|^2
    \\
    & \left.
    +\nu^2 (1-R^{-4})|z|^2 \left(
    |w_+|^2 + |w_-|^2
    + \frac{1}{2} p^2 |\phi|^2
    \right)
    \right\}.
\end{split}
\end{equation}
Due to the gauge constraint (at order $O(\nu)$) $\frac{1}{2}p^2 \phi = D_z w_- + D_{\bar z} w_+$ \eqref{eq:gauge_const}, the action can be further organized into a diagonal form
\begin{equation}
\begin{split}
    I_W^{(2)} = \int d^2z &
    \left\{
    4 |D_{\bar z} w_+|^2 + (p^2 + (R-1/R)^2 + 2\nu)|w_+|^2
    \right. \\
    &
    + 4 |D_z w_-|^2 + (p^2 + (R-1/R)^2 - 2\nu)|w_-|^2
    \\
    &
    + p^2\left(|D_z \phi|^2 + |D_{\bar z} \phi|^2
    + \frac{1}{2}\left(p^2 + (R-1/R)^2\right) |\phi|^2
    \right)
    \\
    & \left.
    +\nu^2 (1-R^{-4})|z|^2 \left(
    |w_+|^2 + |w_-|^2
    + \frac{1}{2} p^2 |\phi|^2
    \right)
    \right\}.
\end{split}
\end{equation}
The fields $w_\pm$ and $\phi$ have an action identical to the transverse mode \eqref{eq:trans_I2}, albeit with a shifted mass. With the same trick, the action can be recast in terms of the Hamiltonian operator \eqref{eq:H_def} as
\begin{equation}
\begin{split}
    I_W^{(2)} = \int d^2z &
    \left\{
    w_+^* \left(\hat H + \nu (1+R^{-2})\right)w_+
    +
    w_-^* \left(\hat H - \nu (1+R^{-2})\right)w_-
    + \frac{1}{2} p^2\phi^* \hat H \phi
    \right\}.
\end{split}
\end{equation}
In this form, the $w_\pm$ spectrum can be read off immediately in the same Landau basis $w_{k,\tilde k}$
\begin{equation}\label{eq:lambda_long}
    \lambda_{k,\tilde k}^{(\pm)}
    = p^2 + (R-1/R)^2 + 2\nu \left( 1 + (1-R^{-2})k + (1+R^{-2})(\tilde k\pm 1) \right) \equiv \lambda_{k,\tilde k \pm 1},
\end{equation}
where in the second equality, we use the notation from the transverse spectrum \eqref{eq:lambda_trans}. This result again matches exactly with the worldsheet analysis \eqref{eq:m2_n_nt}. The shifted $\tilde k$ comes from both the dipole moment coupling and from the kinetic terms.
The W ground state comes from the $w_-=w_{0,0}$, $w_+ = 0$ mode. By the gauge constraint $\frac{1}{2}p^2 \phi = \tilde a w_- - \tilde a^\dagger w_+$, also $\phi=0$. The mass shell of the ground state reads
\begin{equation}\label{eq:W_gs_mass}
    m^2 = (R-1/R)^2 -\frac{2\nu}{R^2} + O(\nu^2).
\end{equation}
This answer coincides with the worldsheet value and yields a target space reasoning for the Hagedorn instability at $\nu>0$.

This concludes our analysis of the W boson spectrum. The W boson effectively experiences a magnetic field $\hat F$, which discretizes the spectrum to Landau levels. Due to the magnetic dipole coupling, the Zeeman splitting allows tachyons for longitudinal polarization.
A second term lifts the original degeneracy of the levels (see appendix \ref{app:adjoint_higgs}) and selects a unique ground state in the W sector, with mass \eqref{eq:W_gs_mass}.

\subsection{Comparison to the electroweak theory}\label{subsec:electroweak}
Our $SU(2)\times U(1)$ theory is reminiscent of another theory with the same gauge group, our very own electroweak theory. Leaving supersymmetry (and gravity!) aside, the main difference is the representation of the Higgs field. Above, the Higgs was an $SU(2)$ triplet, while the Higgs in our world is a complex doublet. This is the reason our theory breaks into $U(1)\times U(1)$ with one complex W boson, while the electroweak theory breaks to a single $U(1)_\text{EM}$ and includes both W and Z vector bosons. Nonetheless, in both theories, W is charged under an unbroken $U(1)$ with the same quadratic theory
\begin{equation}\label{eq:L_mag_1}
    I_W^{(2)} = \int d^d x \left(\frac{1}{4}|(d+i A)\wedge W|^2 + \frac{i}{2} F^{ij} W_{i}W^*_{j} + \frac{1}{2}m^2 |W|^2\right).
\end{equation}
We already saw that for small $\nu$ the background experienced by the W boson is similar to a constant magnetic field. It is thus instructive to recall the phenomenology of a W boson in a constant magnetic field for comparison. See appendix \ref{app:adjoint_higgs} for more technical details.

Under a constant background magnetic field $B>0$, the two-dimensional spectrum of a free particle with spin $j$ and charge $q$ is
\begin{equation}
    E_{k,s}^2 = m^2 + qB(1+2k+2s),
\end{equation}
with $-j\le s \le j$ the spin in the direction of the magnetic field and level $k\ge 0$. For a scalar $j=s=0$, the spectrum is always gapped by the magnetic field, a famous property of type-II superconductors. For our W boson $s=\pm 1$, which gives a Zeeman splitting due to the magnetic dipole coupling in \eqref{eq:L_mag_1}. 
The ground state energy comes from $s=-1$ and $k=0$, and reads
\begin{equation}
    E_{0,-1}^2 = m^2-qB.
\end{equation}
As in \eqref{eq:W_gs_mass} above, high enough magnetic fields turn the W boson into a tachyon. This effect is called the Ambjørn--Olesen phenomenon \cite{Ambjorn:1988tm,Ambjorn:1988fx,Ambjorn:1989bd}.\footnote{
The electroweak theory destabilizes around $B_{c1}^\text{EW} = m^2_{W}/q = 1.1 \cdot 10^{20} \text{ T}$.
However, since in the full standard model the lightest charged vector is not the W boson but the $\rho$ meson, the QCD sector destabilizes into a $\rho$ meson condensation at a lower value $B_c^\text{QCD} = m^2_{\rho}/q = 1.0 \cdot 10^{16} \text{ T}$. For comparison, the highest human-produced magnetic field is $\sim 10^3 \text{ T}$ \cite{nakamura2018record}.}

Landau levels have an enormous degeneracy proportional to the magnetic flux. This degeneracy is crucial to the physics of the electroweak theory above the critical value $B>B_c$.\footnote{At a slightly higher value $B_{c2}^{\text{EW}}>B_c^\text{EW}$ the electroweak symmetry is restored, similar to a type II superconductor \cite{Ambjorn:1989bd}.}
Our background is dramatically different. Since the magnetic Melvin breaks two-dimensional translation invariance, already the quadratic W theory lifts the degeneracy. As a result, slightly below the critical $\nu$, only a single mode $w_- = w_{0,0}$ condenses.

\section{W stars}\label{sec:w_stars}
By tuning the temperature slightly above the Hagedorn temperature $R-R_H \ll 1$, the Atick-Witten scalar can be made arbitrarily light. A self-gravitating normalizable condensate solution of the scalar was found by Horowitz and Polchinski \cite{Horowitz:1997jc}. By expanding in small $R-R_H$, it was possible to consider only the leading Newtonian gravity cubic coupling, which led to a self-consistent background for $2 < d < 6$ spatial dimensions. This background was proposed \cite{Horowitz:1997jc,Chen:2021dsw} as a stringy description of high-temperature black holes.

How would such a solution look like if we were to follow it with $\nu$ close to the index $\nu \ll 1$? Compactifying $9-d$ spatial directions, the partition function \eqref{eq:Z_intro_def} corresponds to a $(d+1)$-dimensional Melvin background. Close enough to $R=1$, the effective theory is the $d$-dimensional reduction of the 9d $SU(2)$ SUGRA theory spelled in the previous section. Instead of a thermal scalar, a winding condensate corresponds to a normalizable saddle for the W boson. Below, we will show how to find such a solution in the $SU(2)$ theory.

At the index, the background is supersymmetric, the Lagrangian is entirely positive, and admits no normalizable saddles.\footnote{Denote such saddle collectively by $\phi(x)$ and consider the variation $\phi(\lambda x)$. For a local $d>2$ dimensional positive action with up to 2-derivatives, it is trivial to show that $\delta_\lambda I < 0$.}
Thus, at $\nu=0$ there are no ``W stars'' saddles. This is a striking difference with the Atick-Witten EFT. The application of this statement for the correspondence with black holes will be discussed below. For now, the lesson is that in order to find such saddles we must turn on $\nu>0$, and shouldn't treat it as a perturbative parameter. Such a saddle can be thought of as a meta-stable bubble of the W condensation phase. The existence of a high-temperature W boson tachyon for $\nu>0$, established in the previous section, is a positive signal that such a bubble solution could be formed.

\paragraph{Terminology} The term ``Horowitz-Polchinski solution" will describe the solution for the thermal-scalar Atick-Witten EFT \cite{Horowitz:1997jc} and its possible generalizations. ``W star'' will describe only W boson normalizable condensate solutions of the Heterotic $SU(2)$ EFT. To describe any normalizable string theory saddle of thermal winding strings, we will use the term ``string stars'' collectively. Thus, both the Horowitz-Polchinski solution and W stars are types of string stars, but not the other way around.

\subsection{Low energy effective action}\label{subsec:low_e_eft}
In order to use the effective action, we tune the temperature so that the winding ground state \eqref{eq:W_gs_mass} is light in string units, $R-R_H(\nu) \ll 1$ and $\nu \ll 1$.
As we will argue below, such tuning will also ensure that the amplitude $W \ll 1$ is small, allowing us to expand to leading order in the fields. At the same time, we will continue to use the leading $O(\nu)$ approximation \eqref{eq:melvin_back_approx}.

Expanding in the fields, the leading interaction would be cubic $\sim W^2 \psi$, where $\psi$ stands for some fluctuation of the background. Consider the background deformations
\begin{equation}
    \sigma = \sigma_0 + \varphi, \quad A_i = A_i^0 + a_i, \quad \bar A_i =  A_i^0 + \bar a_i,
\end{equation}
where the subscript $_0$ stands for the background \eqref{eq:melvin_back}. We didn't include gravity and dilaton deformations, since, as we are about to see, they are subleading.
We start by copying the original $SU(2)$ action from \eqref{eq:W_su2_action}, leaving only the relevant fields $\sigma,A,\bar A ,W$ up to order $W^2$ (and neglecting the quartic $W^4$ term):
\begin{equation}\label{eq:W_su2_action_2}
\begin{split}
    I_{SU(2)} &= \frac{R}{8 G_N} \int d^d x \sqrt{g} e^{-2\phi} \left(\frac{1}{4} (\partial \sigma)^2 
    + \frac{\cosh(\sigma)}{8}\left(F^2+ \bar F^2\right)+ \frac{\sinh(\sigma)}{4} F_{ij} \bar F^{ij}\right.
    \\
    &\left.
    + \frac{1}{4}|DW|^2+ \frac{2\cosh(\sigma)-2}{2} |W|^2
    + \frac{i}{2} \left(\cosh(\sigma) F^{ij}+\sinh(\sigma) \bar F^{ij}\right) W_{i}W^*_{j}
    \right).
\end{split}
\end{equation}
To find the cubic interactions, we expand the action \eqref{eq:W_su2_action_2} to cubic order in $\varphi, a_i, \bar a_i $ and $W$. The zeroth order will be zero since the background is a $10$-dimensional Ricci-flat manifold. The linear order will be canceled by the background equations of motion. We therefore write down only quadratic and cubic terms:
\begin{equation}
\begin{split}
    \frac{1}{4} (\partial \sigma)^2 & =  \frac{1}{4} (\partial \varphi)^2\\
    \frac{\cosh(\sigma)}{8} \left(F^2+\bar F^2\right) +\frac{\sinh(\sigma)}{4} F \bar F
    &=
    \frac{e^{\sigma_0}}{8}(F^0)^2 \varphi^2
    +\frac{e^{\sigma_0}}{4}(\varphi + \frac{1}{2}\varphi^2)F^0_{ij}(f^{ij}+\bar f^{ij})\\
    &\quad +\frac{1}{8}(\cosh(\sigma_0)+\sinh(\sigma_0)\varphi)(f^2+\bar f^2)\\
    &
    \quad +\frac{1}{4}(\sinh(\sigma_0)+\cosh(\sigma_0) \varphi)f \bar f\\
    \frac{1}{4} |DW|^2 &= \frac{1}{4} \left|D_0 W+i a \wedge W\right|^2,\\
    &=\frac{1}{4}|D_0 W|^2 + \frac{i}{2} a_i \left((D_0W^*)^{ij} W_j -W_j^* (D_0W)^{ij}\right),\\
    \frac{2\cosh(\sigma)-2}{2} |W|^2 &= \frac{2\cosh(\sigma_0)-2}{2} |W|^2 + \sinh(\sigma_0) |W|^2\varphi\\
    \frac{i}{2} \left(\cosh(\sigma) F^{ij}+\sinh(\sigma) \bar F^{ij}\right)
    W_{i}W^*_{j} &=
    \frac{i}{2}e^{\sigma_0} (F^0)^{ij}W_{i}W^*_{j}
    +\frac{i}{2}e^{\sigma_0} (F^0)^{ij} W_{i}W^*_{j} \varphi\\
    &+\frac{i}{2} (\cosh(\sigma_0) f^{ij} + \sinh(\sigma_0) \bar f^{ij}) W_{i}W^*_{j},
\end{split}
\end{equation}
with $D_0 W= dW + i A^0 \wedge W$.
Using \eqref{eq:melvin_back_approx}, we expand to order $O(\nu)$, and assume $\partial \sim \nu^{1/2}$. 
To leading order, the quadratic piece is\footnote{We expanded the first line up to order $\nu^0$, but the second line to $\nu^1$. The reason is that the latter is actually leading in $R-R_H(\nu)$.}
\begin{equation}
\begin{split}
    I_2 = \frac{R}{8 G_N} I_{W}^{(2)} + \frac{R}{8 G_N} \int  &d^{d} x\left(\frac{1}{4} (\partial \varphi)^2
    + \frac{R^2+R^{-2}}{16}\left(f^2+ \bar f^2\right)+ \frac{R^2-R^{-2}}{8} f_{ij} \bar f^{ij}
    \right),
\end{split}
\end{equation}
with
\begin{equation}
    I_{W}^{(2)} = \int  d^{d} x\left(
    \frac{1}{4}|D_0 W|^2+ \frac{(R-1/R)^2 + \nu^2 (1-R^{-4})|z|^2}{2} |W|^2
    + i\nu (W_1 W^*_2-W_2 W^*_1)
    \right),
\end{equation}
the quadratic W action \eqref{eq:pert_action_2} we analyzed in the previous section. Up to $O(\nu^2)$ cubic couplings the interactions are
\begin{equation}
\begin{split}
    I_3 &= \frac{R}{8 G_N} \int  d^{d} x\left(
    \frac{R^2-R^{-2}}{2}\varphi |W|^2
    +
    \frac{i}{2} a_i \left((D_0 W^*)^{ij} W_j - c.c.\right)
    \right.
    \\
    &
    + \frac{i}{2}(R^2+R^{-2})f^{ij} W_i W_j^*
    + \frac{i}{2}(R^2-R^{-2})\bar f^{ij} W_i W_j^*
    \\
    &\left.
    +\frac{R^2-R^{-2}}{16} \varphi (f^2+\bar f^2) + \frac{R^2+R^{-2}}{8}\varphi f \bar f
    + i \nu \varphi (W_1 W^*_2-W_2 W^*_1)
    \right).
\end{split}
\end{equation}
However, by derivative suppression ($\partial \sim \nu^{1/2}$), all the 2-derivative terms in the third line are actually subleading. This is also the reason we didn't write down the dilaton and the metric cubic couplings. The dilaton and the metric will couple to a quadratic term in the action, all of which are either of order $\nu^1$ (by derivative counting) or $R-R_H$, both subleading couplings.

Finally, to leading order in $R-R_H(\nu)$, we can set $R=R_H(\nu)$ in the cubic couplings. Since 
\begin{equation}
    R^2_H(\nu)-R^{-2}_H(\nu) = 2\sqrt{2\nu}+O(\nu),
\end{equation}
the quadratic $f_{ij} \bar f^{ij}$ and the cubic $\bar f^{ij} W_i W_j^*$ terms are of subleading order, which set $\bar a_i = 0$ at this order.
Altogether, we end up with the cubic action
\begin{equation}
\begin{split}
    I_\text{eff} = \frac{R}{8 G_N} I_{W}^{(2)} + \frac{1}{8G_N} \int d^{d} x \left(
    \frac{1}{4}(\partial \varphi)^2 + \frac{1}{8} f^2
    +\sqrt{2\nu} \left(|W|^2 \varphi + a_j J^j\right)
    \right),
\end{split}
\end{equation}
with
\begin{equation}\label{eq:j_def}
    J^j = \frac{i}{2\sqrt{2\nu}} \left((D_0W^*)^{ji}W^i-(D_0W)^{ji} W^{*,i}-\partial_i(W^{[i}W^{*,j]})\right),
\end{equation}
is the electric current, normalized so that by derivative counting $J \sim \nu^0$.
This form of the action is already quite similar to Horowitz and Polchinski's. The two differences are that the winding mode is a vector and not a scalar, and that the leading cubic coupling involves both the Newtonian potential $\varphi$ and the KK vector $a_i$ (this is similar to \cite{Seitz:2025wpc}).

The action for both $\varphi$ and $a_i$ is easy to integrate out. In Lorentz gauge $\partial_i a^i = 0$, their equations of motion are
\begin{equation}
\begin{split}
    \nabla^2 \varphi &= 2\sqrt{2\nu} |W|^2,\\
    \nabla^2 A^j &= 2\sqrt{2\nu} J^j
\end{split}
\end{equation}
Substituting the solution
\begin{equation}
\begin{split}
    \varphi(x) &= -2\sqrt{2\nu} C_{d} \int d^{d} x' \frac{|W|^2(x')}{|x-x'|^{d-2}},\\
    a_i(x) &= -2\sqrt{2\nu} C_{d} \int d^{d} x' \frac{J_i(x')}{|x-x'|^{d-2}},
\end{split}
\end{equation}
with $C_d = \frac{\Gamma\left(\frac{d-2}{2}\right)}{4\pi^{d/2}}$, the on-shell action is
\begin{equation}\label{eq:eft_w}
\begin{split}
    I_\text{eff} &= \frac{R}{8 G_N} \left\{
    I_{W}^{(2)} -2\nu \ C_{d} \int d^{d} x \ d^{d} x'
    \frac{|W(x)|^2 |W(x')|^2 +J^i(x) J_i(x')}{|x-x'|^{d-2}}
    \right\}
    .
\end{split}
\end{equation}
The double integral introduces the two long-range forces that govern the dynamics.
The first term is negative definite and corresponds to (Newton's) gravitational attraction. The second term is Ampère's magnetic force between electric currents $J^i$. This force is attractive at short distances when the currents are aligned, but otherwise has an undetermined sign. To find a solution, the overall self-force must be sufficiently attractive.
Moreover, local quadratic terms were neglected in the derivation of \eqref{eq:eft_w}.
Due to supersymmetry, these terms are positive (repulsive force), and could theoretically win over the attractive long-range force. As we will see below, provided the solution's size is sufficiently large, the repulsive terms remain subleading.

\subsection{An even-lower energies EFT}\label{subsec:lower_E_EFT}
So far, we have only expanded the effective action to leading order in both $R-R_H(\nu)$ and $\nu$. It is theoretically possible to seek numerical solutions of the EFT. There is, however, a further simplification one can make.
For a fixed $\nu$, the two-dimensional winding spectrum is discrete with spacing $\Delta m^2 \sim \nu$. It is possible to tune the temperature so that the mass of the winding ground state \eqref{eq:W_gs_mass} is arbitrarily light compared to the rest of the discrete spectrum, namely
\begin{equation}\label{eq:low_e_limit}
    m^2 \ll \Delta m^2 \ll 1,
\end{equation}
or $R-R_H(\nu) \ll \sqrt{\nu}$.\footnote{In fact, when $R-R_H(\nu) \ll \sqrt{\nu}$, there are chiral $\tilde k =0$ modes with softer spacing \eqref{eq:lambda_long} of $\Delta m^2 \sim \nu^{3/2}$. Since the solution we are looking for has zero orbital angular momentum $l=k-\tilde k=0$, we can ignore these modes. Considering these modes leads to a tighter bound of $R-R_H \ll \nu$.} Assuming the solution size is much larger than the gap, the only winding modes in this limit are light $(d-2)$-dimensional momentum modes of the W ground state.

For the rest of the section, we use $z,\bar z$ for the two-dimensional rotation plane, while $x^i$ stands for the rest of the coordinates $i=3,...,d$. In these conventions, our limit allows us to set 
\begin{equation}\label{eq:W_ansatz}
    W = W_0(z,\bar z) \cdot \chi(x), \quad 
    W_0(z,\bar z) = \sqrt{\frac{\nu}{\pi}} \exp(-\nu |z|^2/2) d\bar z.
\end{equation}
The ground state $W_0$ \eqref{eq:w_gs_wavefunction_nu} is a two-dimensional vortex solution of size $|z|^2 \sim 1/\nu$. The normalization of $W_0$ was chosen so that $\int d^2 z |W_0|^2 = 1$.
The vortex's mass and current densities \eqref{eq:j_def} are 
\begin{equation}
    |W_0|^2 = \frac{\nu}{\pi} \exp(-\nu |z|^2), \quad 
    J = |W_0|^2 \cdot \sqrt{\frac{9\nu}{2}} \frac{z d\bar z - \bar z dz}{2i}.
\end{equation}
Substituting in \eqref{eq:eft_w} leads to an effective $(d-2)$-dimensional theory for $\chi$
\begin{equation}\label{eq:eft_w_2}
\begin{split}
    I_\text{eff} = \frac{R}{8G_N}\int d^{d-2} x &\left\{
    \frac{1}{2} |\partial \chi|^2 + \frac{1}{2} m^2 |\chi|^2 
    -2\nu \int d^{d-2} x' G(|x-x'|) |\chi(x)|^2 |\chi(x')|^2\right\},
\end{split}
\end{equation}
with the kernel ($r=|x|$)
\begin{equation}
\begin{split}
    G(r) &=  C_{d} \int d^2 z d^2 z'
    \frac{|W_0(z,\bar z)|^2 |W_0(z',\bar z')|^2 +J_0(z,\bar z) \cdot J_{0}(z',\bar z')}{\left(r^2 + |z-z'|^2 \right)^\frac{d-2}{2}}\\
    &= C_{d} \frac{\nu^2}{\pi^2} \int d^2 z d^2 z' e^{-\nu (|z|^2+|z'|^2)}
    \frac{1 +\frac{9 \nu}{4} (z \bar z'+\bar z z')}{\left(r^2 + |z-z'|^2 \right)^\frac{d-2}{2}}.
\end{split}
\end{equation}
To get \eqref{eq:eft_w_2}, we neglected the $(d-2)$-dimensional component of the current density since our solution for $\chi$ would be real. Numerically, we found that both terms in the numerator integrate to a positive value (although the integrand does not) $G>0$, so that the kernel always induces an attractive force.

The $(d-2)$-dimensional effective theory is similar in spirit to the Horowitz-Polchinski one, in the sense that both have a single scalar field $\chi(x)$ with some long-range self-force. Yet, since the effective theory here is $(d-2)$-dimensional, the long-range gravitational force is now marginal.
At large $(d-2)$-dimensional distances $\nu r^2 \gg 1$, the kernel is dominated by the Newtonian fall off $G(r) \sim C_{d}/r^d$. At $m^2 \ll \nu$, it is tempting to assume that the solution is large and solve the system with a purely Newtonian potential. Such potential, however, yields marginal logarithmic UV divergences in \eqref{eq:eft_w_2}. Thus, we must keep the explicit kernel dependence on the `UV scale' $\nu$, schematically given by ($c >0$ is an $O(1)$ number)
\begin{equation}\label{eq:G_approx}
    G(r) \approx \frac{C_d}{(r^2 + c /\nu)^{\frac{d-2}{2}}},
\end{equation}
due to the $|z|^2 \sim \nu^{-1}$ spread of $W_0$. The UV scale smooths out the gravitational power-law behavior at short distances $\nu r^2 \lesssim 1$ to $G(r) \sim \nu^{\frac{d-2}{2}}$, which will regulate the marginal divergences.

To get a better picture, let us use the variational method, and set an ansatz of the type 
\begin{equation}\label{eq:ansatz_chi}
    \chi(x) = \chi_0 \cdot f(|x|/L),
\end{equation}
with $f$ a normalizable function with no intrinsic scale, amplitude $\chi_0$ and size $L$. As a concrete example, one can choose $f(x) = \exp(-x^2/2)$
together with the kernel \eqref{eq:G_approx}. Our analysis would use a general $f$, which would presumably hold also for the real saddle of the action.
The action for the ansatz \eqref{eq:ansatz_chi} can be written as
\begin{equation}\label{eq:I_ansatz}
\begin{split}
    I_\text{eff} = \chi_0^2 \ L^{d-2} (L^{-2} I_1 + m^2 I_2)
    -\nu \ \chi_0^4\ L^{d-2}\left(I_3 +I_4 \log(L^2 \nu)\right)
\end{split}
\end{equation}
with $I_1,...,I_4$ all positive $O(1)$ numbers that depends solely on $f$. In the last term, we took only the leading $L^2 \nu \gg 1$ limit. Extermizing the action for $\chi_0$ gives
\begin{equation}\label{eq:A}
    \chi_0^2 = \frac{1}{2L^2 \nu} \frac{I_1 + m^2 L^2 I_2}{I_3 + I_4 \log(L^2 \nu)}.
\end{equation}
Substituting \eqref{eq:A} in the equation for $L$ gives
\begin{equation}
    \frac{\partial I}{\partial L}= \frac{\chi_0^2 L^{d-5}}{2} \left((d-6) I_1 + (d-2)\ I_2 \ m^2 L^2
    -\frac{2I_4(I_1+m^2 L^2 I_2)}{I_3 + I_4 \log(L^2\nu)}\right)=0.
\end{equation}
Assuming that the solution is large compared to the UV scale $L^2 \nu \gg 1$, the answer is approximately $m^2 L^2 \approx \frac{6-d}{d-2} \frac{I_1}{I_2}$ and, by \eqref{eq:A},
\begin{equation}
    \chi_0^2 \approx \frac{m^2}{\nu} \frac{2 I_2}{6-d} \frac{1}{I_3 + I_4 \log\left(\frac{6-d}{d-2}\frac{I_1}{I_2} \nu/m^2\right)}.
\end{equation}
To maintain $L^2 >0$, a solution exists only for $2<d<6$.
In the low energy limit $m^2 \ll \nu$ the answer we found scales as
\begin{equation}\label{eq:I_sol}
    \chi_0^2 \sim \frac{m^2}{\nu} \frac{1}{\log(\nu/m^2)},\quad L \sim 1/m.
\end{equation}
and thus self-consistently satisfies both $\chi_0\ll 1$ and $L^2 \nu \gg 1$. The system \eqref{eq:eft_w_2} can also be solved directly using numerical methods. Using the approximated kernel \eqref{eq:G_approx} we were able to find normalizable solutions for $\chi$ that matched (for small $m^2 \ll \nu$) the variational results.

Let us go back and address the issue with the quartic coupling. The term we neglected from \eqref{eq:W_su2_action} was $[W,W]^2$, which translates in \eqref{eq:I_ansatz} to a term $+\nu \chi_0^4$. This term could shift the $I_3$ term to negative, but it remains subdominant due to the logarithmic enhancement of the coupling. Whenever the logarithmic enhancement is no longer dominant, we no longer have $L^2 \nu \gg 1$, and one needs to retreat to the full $d$-dimensional theory. Thus, at least for the purpose of the $d-2$ dimensional calculation, the quartic coupling was consistently ignored.
\newline

\subsection{Properties of the W star}\label{sec:W_star_prop}
Using \eqref{eq:I_sol}, to leading order in small $m^2$, the on-shell action, energy, and entropy of the solution are
\begin{equation}\label{eq:W_I_E}
    I \sim \frac{1}{G_N} \frac{m^{6-d}}{\nu \log(\nu/m^2)}, \quad
    E = \frac{1}{2\pi} \partial_R I = \frac{4\sqrt{2\nu}}{16\pi G_N} \int d^{d-2} x |\chi|^2 \sim \frac{1}{G_N}\frac{m^{4-d}}{\nu^{1/2} \log(\nu/m^2)}.
\end{equation}
In this language, we also define 
\begin{equation}\label{eq:S_nu_def}
    S_\nu \equiv S + \beta \Omega J \equiv (R\partial_R -1)I.
\end{equation}
For $\nu=1$ it is simply the entropy, while at $\nu=0$ it gives the index $S_\text{index} = S + 2\pi i J$.
Adding string units for clarity,
\begin{equation}
\label{eq:hag_w_star}
     S_\nu = 2\pi l_s E.
\end{equation}
The relation \eqref{eq:hag_w_star} is interesting because it holds the Hagedorn growth of the system close to the index at which ``$\beta_H = 2\pi l_s$'', although at the index itself no Hagedorn temperature exists. 

A transition to the free-string phase is expected to occur when the on-shell action is so small that quantum effects begin to dominate \cite{Chen:2021dsw}. Setting $I \sim 1$ gives $m^{6-d}\sim \nu g_s^2$, where $g_s$ is the $D$-dimensional string coupling. Therefore, the temperature regime of validity is 
\begin{equation}\label{eq:R_validity_regime}
    (R-R_H)/l_s \gg \nu^\frac{d-2}{2(6-d)} g_s^\frac{4}{6-d}.
\end{equation}
At the transition, the energy and entropy are of order
\begin{equation}\label{eq:free_string_trans}
    l_s E, S \sim g_s^{-\frac{4}{6-d}} \nu^{-\frac{d-2}{2(6-d)}}.
\end{equation}

The properties of the W star are reminiscent of the Horowitz-Polchinski solution \cite{Horowitz:1997jc,Chen:2021dsw}. 
There as well, a solution was found only for $2<d<6$ spatial dimensions. In terms of the thermal scalar mass $m^2$, the Horowitz-Polchinski solution has a similar size $L \sim 1/m$, amplitude $\chi_0 \sim m$, and action $I \sim m^{6-d}$ scalings, only without the logarithmic suppression. Setting $\nu \sim 1$ to obtain the thermal case yields the correct behavior. Thus, it is very likely that the Horowitz-Polchinski solution and the W star live along the same line parameterized by $\nu$.
To see why, recall that for any $0
<\nu<1$ the winding spectrum is discrete (see figure \ref{fig:heterotic_spectrum}), and we can take the low energy limit \eqref{eq:low_e_limit}. Upon integrating the long-range forces, we will end up with something similar to \eqref{eq:eft_w_2}, with $\nu$ as a continuous parameter. In this description, it is clear that as long as the effective force remains attractive, a string star solution will interpolate the W star to the Horowitz-Polchinski solution. See figure \ref{fig:transition}
\newline

\begin{figure}
    \centering
    \includegraphics[width=0.7\linewidth]{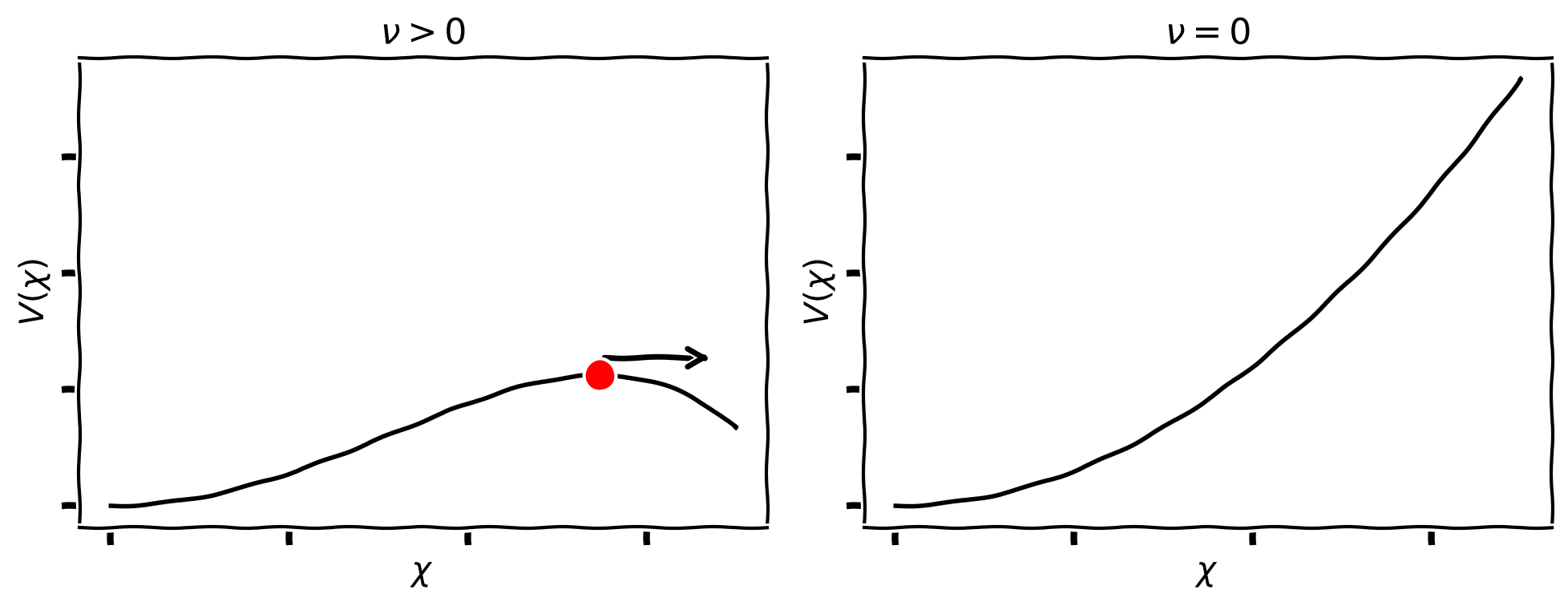}
    \caption{A schematic picture of the the winding amplitude $\chi$ effective potential. On the left at $\nu > 0$, a non-trivial meta-stable ``W star'' solution exists around $\chi^2 \sim 1/\nu$. As we take the limit $\nu=0$, the solution grows and eventually decouples. At $\nu=0$, on the right panel, the winding mode is stable and admits no non-trivial solutions.}
    \label{fig:potential}
\end{figure}

As explained above, no W star solutions exist at $\nu=0$. Nonetheless, we can still ask what happens to the solutions in the $\nu=0$ limit.
So far, we have found solutions at small but fixed $\nu$ and temperatures arbitrarily close to the Hagedorn temperature $m^2 \ll \nu$. Since this regime is empty at $\nu=0$, we can't use the results to answer this question. Let us work in terms of a fixed winding mass $m^2$. Although in principle higher Landau levels can participate in the W star profile, since only the ground state is tachyonic, \eqref{eq:W_ansatz} should give a good estimate for the profile. Since the perpendicular modes are now highly coupled to the Landau levels, we estimate the W star's size in this regime to be  $L\sim 1/\sqrt{\nu}$.
We also expect the on-shell action $I \sim 1/\nu$ scaling in \eqref{eq:W_I_E} to persist at fixed $m$. In the $d$-dimensional EFT for $W$ \eqref{eq:eft_w}, the attractive quartic coupling scales like $\nu$. Schematically, we have an effective potential for the winding amplitude $\chi_0$
\begin{equation}\label{eq:V_eff}
    V_\text{eff} = m^2 \chi_0^2 - \nu \chi_0^4 + ....
\end{equation}
See figure \ref{fig:potential} for an illustration.
This type of potential is inline with the schematic form we found above \eqref{eq:eft_w}. It admits a meta-stable solution at $\chi_0^2 \sim m^2/\nu$ with the same $V \sim 1/\nu$ scaling. Therefore, in the fixed temperature $\nu=0$ limit, the W star doesn't dissolve, but decouples with an arbitrarily large free energy. 
Moreover, since the amplitude also grows in the limit, it would be impossible to reliably find saddles in the limit, unless we recede to the $m^2 \ll \nu$ limit.

Instead of fixing the temperature, one can fix the energy in the $\nu=0$ limit. The temperature dependence of the energy \eqref{eq:W_I_E} is determined by the sign of $d-4$. Yet it seems likely that, for any $d$, the W star (just like the Horowitz-Polchinski solutions) transitions at high temperatures to the lower-energy free-string phase \cite{Chen:2021dsw}.  The transition energy scale \eqref{eq:free_string_trans} goes to infinity at $\nu=0$. Thus, the simplest interpretation would be that the W star decouples on the $\nu=0$ limit, and transitions to the free string phase.

\subsection{The black hole/ string transition}\label{subsec:bh_w_trans}
\begin{figure}
    \centering
    \includegraphics[width=0.5\linewidth]{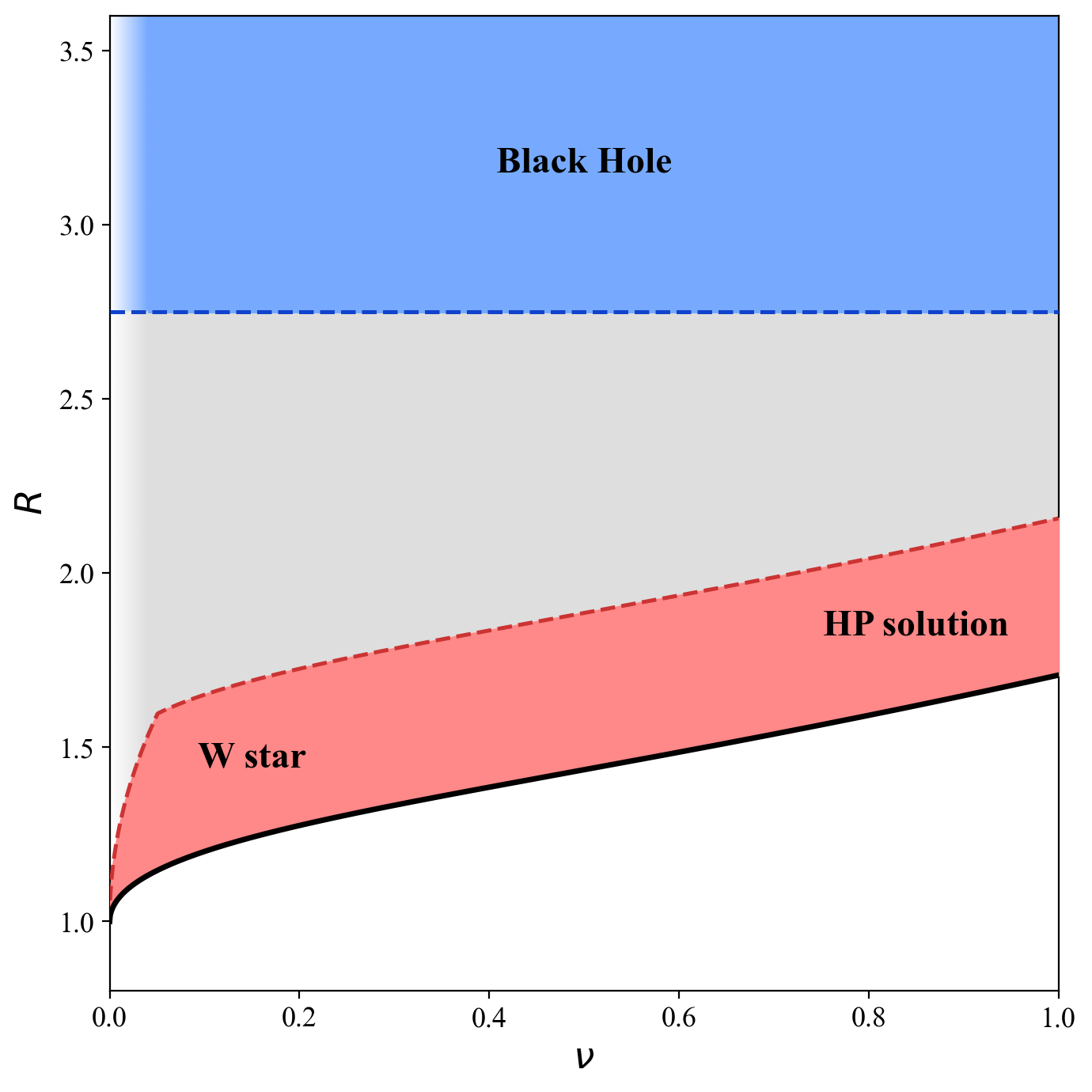}
    \caption{
    The different regimes of validity of the black hole (blue), the Horowitz-Polchinski solution (red, right), and the W star (red, left) in the $R,\nu$ parameter space. The thick black line is the Hagedorn temperature $R_H(\nu)$.
    Rotating black holes are weakly curved for $R\gg 1$, and exist for $\nu>0$. The Horowitz-Polchinski solution is a reliable solution of the Atick-Witten EFT for $|1-\nu| \ll 1$ and $R-R_H \ll 1$. The W star is a reliable solution of the $SU(2)$ EFT for $|\nu| \ll 1$ and $R-1 \ll \sqrt{\nu}$.
    At $\nu=0$ both the W star and the weakly-curved black hole decouple.
    }
    \label{fig:transition}
\end{figure}

The Horowitz-Polchinski solution was suggested as a possible description of black holes with temperature $R \sim 1$ (in string units). In \cite{Horowitz:1997jc,Chen:2021dsw} a rough agreement between the thermodynamic properties of the two was found in an intermediate regime of temperatures and energies.
In \cite{Seitz:2025wpc} it was extended for small zero angular velocity ($\beta \Omega 1$).
The W star is the continuous deformation of the Horowitz-Polchinski solution to small $\nu \sim 0$. 
A non-trivial check of the string-black hole correspondence would be to compare the W star properties we spelled in the previous section, to those of a rotating black hole close to $\nu = 0$.

We will follow the conventions of~\cite{Emparan:2003sy,Ceplak:2023afb}.
The thermodynamic properties of rotating black holes in $D=d+1$ dimensions (also known as Myers-Perry black holes~\cite{Myers:1986un}) are best described using the variables $m$, $a$, and the black hole horizon radius $r_h$. Since the solution is purely Euclidean, we set $a=ib$. These variables 
are related to the mass, spin, and entropy of the black hole by
\begin{equation}\label{eq:MSJ}
\begin{split}
    M = \frac{\Omega_{D-2}}{8\pi G_N} (D-2) m, \quad S = \frac{\Omega_{D-2}}{2 G_N} m r_h, \quad
    J = i \frac{\Omega_{D-2}}{4\pi G_N} m b,
\end{split}
\end{equation}
with
\begin{equation}
    r_h^2-b^2 = \frac{2m}{r_h^{D-5}}.
\end{equation}
Note that for both the entropy and the mass to be positive, we need $r_h > |b|$. We will assume $b>0$ from now on.

Defining $R = \beta/(2\pi)$, the temperature, angular velocity and the twist $\beta \Omega = 2\pi i (1-\nu)$ are parametrised as
\begin{equation}\label{eq:R_bh}
\begin{split}
    R = 
    \frac{2(r_h^2-b^2)r_h}{(D-3)r_h^2+(5-D)b^2}, \quad
    \Omega = i \frac{b}{r_h^2-b^2},
\end{split}
\end{equation}
\begin{equation}\label{eq:nu_def_bh}
    \nu = 1-\frac{2 b\  r_h}{(D-3) r_h^2 + (5-D) b^2}.
\end{equation}
By \eqref{eq:S_nu_def}, the Smarr formula is
\begin{equation}\label{eq:S_nu}
\begin{split}
    \frac{S_\nu}{2\pi R} & = \frac{\Omega_{D-2} m}{4 \pi G_N} \frac{r_h-b}{R} = \frac{D-3}{D-2}\  M.
\end{split}
\end{equation}

We are interested in the smooth extrapolation of the non-rotating solution (at $\nu=1$) to $\nu=0$.
As we will see shortly, this is a singular limit. One way to take the limit is at fixed temperature $R$.
Using \eqref{eq:R_bh},\eqref{eq:nu_def_bh}, we write down $r_h$, $b$ and $m$ in terms of $R$ and $\nu$ 
\begin{equation}
\begin{split}
    r_h &= \frac{D-4+\sqrt{1+(D-3)(D-5)(1-\nu)^2}}{2(2-\nu)\nu} R,\\
    b &= \frac{-1+(D-3)(1-\nu)^2+\sqrt{1+(D-3)(D-5)(1-\nu)^2}}{2(2-\nu)(1-\nu)\nu} R.
\end{split}
\end{equation}
For $D>4$ and around $\nu \sim 0$ 
\begin{equation}\label{eq:D_ge_4_nu0}
    r_h,b \sim \frac{D-4}{2} \frac{R}{\nu},\quad\text{and}\quad 
    m \sim \frac{1}{2}\left(\frac{D-4}{2} \frac{1}{\nu}\right)^{D-4} R^{D-3}.
\end{equation}
The fixed-temperature $\nu=0$ limit is therefore singular, yielding a diverging mass, entropy (area), and angular momentum!
Conversely for $D=4$ \cite{Chen:2023mbc},
\begin{equation}\label{eq:D_eq_4_nu0}
    r_h, b \sim \frac{R}{2\sqrt{2\nu}}, \quad\text{and}\quad 2m \sim R.
\end{equation}
This time, the mass $M$ and $S_\nu$ are finite in the limit (!), while the area and the spin are still diverging. The fact that the mass is finite requires further investigation. For example, one can embed the black hole in AdS$_4$. As we approach $\nu=0$, the solution grows and begins to sense the AdS geometry, leading to an increase in mass.
As in the thermal case, AdS black holes have a minimal temperature $R_\text{max} \sim l_{ads}\cdot\nu$. Thus, for low enough $\nu$, the fixed $R$ solution simply disappears.
In any case, for any $D \ge 4$, the horizon radius $r_h$ diverges at $\nu = 0$, which is by itself an alarming sign.

Instead of the temperature, we could also fix the mass as we take the limit. For any $D\ge 4$, we again find that the horizon radius diverges, and for $D>4$, by \eqref{eq:D_ge_4_nu0}, we also reach infinite temperature  $R\rightarrow 0$. Since we trust the EFT only away from the string scale $R\gg l_s$, there are no reliable solutions in such a limit.\footnote{This is not entirely obvious. For small $\nu$, the highest curvature of the solution is at the ``poles'', at which the curvature square $R_{ijkl}R^{ijkl}\sim R^{-4}$, independent of $\nu$.} 
\newline

We are now ready to compare the black hole to the W star at fixed $\nu$, by extrapolating both saddles to an intermediate regime of temperatures. Bringing back $l_s$, and
extrapolating the black hole results to $R \sim l_s$ gives mass, entropy, and horizon radius
\begin{equation}
    M, S_\nu \sim \frac{l_s^{D-2}}{G_N} \frac{1}{\nu^{D-4}},\quad r_h \sim l_s/\nu,
\end{equation}
for $D>4$, and
and
\begin{equation}
    M \sim \frac{l_s^{D-2}}{G_N}, \quad S_\nu \sim \frac{l_s^{D-2}}{G_N} \frac{1}{\nu^{1/2}},
    ,\quad r_h \sim \frac{l_s}{\nu^{1/2}},
\end{equation}
at $D=4$.
Conversely, extrapolating the W star results \eqref{eq:W_I_E}, \eqref{eq:hag_w_star} to $\alpha' m^2 \sim 1$ leads to
\begin{equation}\label{eq:W_low_temp}
    M, S_\nu \sim \frac{l_s^{D-2}}{G_N} \frac{1}{\nu^{1/2}} 
    , \quad L \sim \frac{l_s}{\nu^{1/2}}.
\end{equation}
The estimation of W star's size assumes, as we argued above, that the ground-state Landau level sets the overall scale.
Up to the precise divergence in $\nu$, both saddles agree for $D>4$. The $D=4$ situation is, in general, more elusive, and it's possible that higher derivative terms render a $\nu=0$ divergence in the black hole mass. 

Importantly, both the (weakly curved) black hole and the W star saddles decouple in the $\nu=0$ limit. While each saddle is reliable in some regime of temperatures at any fixed $\nu \ne 0$, both regimes disappear from the spectrum in the limit. Both turn arbitrarily large and (at least for $D>4$) heavy.
More precisely, the W star $\nu=0$ limit is similar to its low temperature limit, in the sense that in both the amplitude becomes large. Since at low temperatures the W star arguably transitions into a stringy black hole, a similar situation could also hold for the W star $\nu=0$ limit (before decoupling as well). For an illustration of the suggested picture, see figure \ref{fig:transition}.

\section{Charged saddles}\label{sec:charged_saddles}
\subsection{Setup}\label{subsec:charged_setup}
In order to consider a non-trivial index, we need to turn on charges. Let one of the $10-D$ compatified directions be an $S^1_r$ with circumference $2\pi r$. It is then possible to define the trace \eqref{eq:Z_intro_def} over the Hilbert space of fixed momentum and winding charges $Q_n$, $Q_w$. At $\nu=0$, the partition function computes an index and is thus independent of the temperature and string coupling. In the free string limit $g_s=0$, the index was computed with entropy \cite{Dabholkar:1989jt,Dabholkar:1990yf}
 \begin{equation}\label{eq:S_micro}
     S_\text{micro} = 4\pi \sqrt{|Q_n Q_w|},
 \end{equation}
for large $Q_n,Q_w \gg 1$.\footnote{Strictly speaking the trace \eqref{eq:Z_intro_def} vanishes at $\nu=0$. In order to get the number of $\frac{1}{2}$-BPS multiplets, it is necessary to take four derivatives of $\nu$ \cite{Chowdhury:2024ngg}. This object is called the fourth helicity supertrace \cite{Kiritsis:1997gu}. This will not affect our work, since we only consider the leading exponential behavior.}
The solutions we found in the previous section are all neutral. 
To find charged solutions, we follow the solution generation technique presented in \cite{Chen:2021dsw}. Below, we will describe the charged black holes and W stars that contribute to the trace at fixed $\nu$, and how they approach the index at $\nu=0$.

Since the same method can be used for both the black hole and the W star saddles, we begin by reviewing it first. For full derivation, see appendix \ref{app:sol_gen}.
Upon reduction to $d$ dimensions, the low-energy effective action has a classical $O(10-d,26-d; \mathbb{R})$ symmetry, with an $O(2,2;\mathbb{R})$ subgroup that acts on the asymptotic $S^1_R\times S^1_r$ geometry.\footnote{The symmetry is believed to hold to any $\alpha'$ order. Quantum-mechanically, only the subgroup $O(2,2;\mathbb{Z})$ remains a symmetry (duality).}
With an appropriate choice of an $\Omega \in O(2,2;\mathbb{R})$ group element, it is possible to transform a neutral saddle into a saddle that carries chosen $S^1_r$ charges. For consistency, the transformation $\Omega$ is chosen so that it maps the winding mode charge vector to itself. Thus, the transformation implicitly depends on the thermal momentum $n$, which itself depend on $\nu$ by
\cite{Seitz:2025wpc}
\begin{equation}\label{eq:n_val}
    n = \begin{cases}
        0 & \text{Bosonic, type II,}\\
        \frac{2-2\nu + \nu^2}{2} & \text{Heterotic}
    \end{cases}.
\end{equation}

As in \cite{Chen:2021dsw}, the parameters of the seed (neutral) solution are denoted with a tilde superscript.
Being a classical symmetry, the tree-level partition functions before and after the transformation are the same:
\begin{equation}\label{eq:Z_equality}
    \log Z(R,r,...) = \log Z(\tilde R , \tilde r,...).
\end{equation}
The parameter $\nu$ is fixed by the transformations and remains the same on both sides of \eqref{eq:Z_equality}.\footnote{$\nu$ can be absorbed into the definition of the coordinates, as in \eqref{eq:flat_metric}. Since the $O(2,2;\mathbb{R})$ transformation acts locally on the fields, it doesn't touch the definition of $\nu$.} 
Using the canonical relation
\begin{equation}
    \log Z = S_\nu -2\pi R \left(M-\mu_n Q_n -\mu_w Q_w\right),
\end{equation}
and $S_\nu$ defined in \eqref{eq:S_nu_def},
one can express the thermodynamic properties of the new solution in terms of the original one by taking derivatives of both sides in \eqref{eq:Z_equality}.
The result is
\begin{equation}\label{eq:map_vars}
\begin{split}
    R &= \frac{\tilde R+n/\tilde R}{2}\cosh \upalpha  +\frac{\tilde R-n/\tilde R}{2}\cosh \upbeta ,\\
    r &= \frac{\tilde R (1+\cosh(\upalpha-\upbeta))}{(\tilde R+n/\tilde R)\cosh \upalpha  + (\tilde R-n/\tilde R)\cosh \upbeta } \tilde r,\\
    Q_L &= e^{-2\phi_D} \frac{\tilde S_\nu'}{2\pi \tilde R}\sinh \upalpha \cosh \upbeta , \quad 
    Q_R = e^{-2\phi_D} \frac{\tilde S_\nu'}{2\pi \tilde R} \cosh \upalpha \sinh \upbeta, \\
    S_\nu & = e^{-2\phi_D} \tilde S_\nu' \left(
    \frac{1-n/\tilde R^2}{2}\cosh \upalpha  +\frac{1+n/\tilde R^2}{2}\cosh \upbeta 
    \right),\\
    M &= e^{-2\phi_D} \left( \tilde M' + \frac{\tilde S_\nu'}{2\pi \tilde R} \left(\cosh \upalpha \cosh \upbeta -1\right)\right),
\end{split}
\end{equation}
with
\begin{equation}
    Q_L,Q_R = Q_n/r \pm Q_w r,
\end{equation}
and $\upalpha,\upbeta$ parametrize the transformation $\Omega$. Alternatively, $\upalpha,\upbeta$ are determined by the charges $Q_L$,$Q_R$. The prime stands for a normalization of the extensive parameters stripped from the seed's asymptotic dilaton value
\begin{equation}
    \tilde M' \equiv e^{2\tilde\phi_D} \tilde M, \quad \tilde S'_\nu \equiv e^{2\tilde\phi_D} \tilde S_\nu.
\end{equation}
This allows us to write down \eqref{eq:map_vars} in terms of the transformed dilaton, given by
\begin{equation}\label{eq:phi_D_trans}
    e^{2\phi_D} =  \frac{R}{\tilde R} e^{2\tilde \phi_D}.
\end{equation}

We stress that, as in \cite{Chen:2021dsw}, \eqref{eq:map_vars} is $\alpha'$-exact. In \cite{Chen:2021dsw}, the transformation was presented separately for type II/Bosonic and Heterotic strings due to the different thermal momenta $n$. 
The relation \eqref{eq:map_vars} is a generalization of \cite{Chen:2021dsw} for general $n$, and $\nu$.

\subsection{Charged black holes}\label{subsec:charged_bh}
Let the seed solution be a weakly curved $\tilde R \gg 1$ neutral Euclidean black hole. Expanding \eqref{eq:map_vars} to leading order in $\tilde R$, and using the Smarr formula \eqref{eq:S_nu}
\begin{equation}\label{eq:map_vars_bh}
\begin{split}
    R = \frac{\cosh \upalpha  +\cosh \upbeta }{2} \tilde R, &\quad r = \frac{1+\cosh(\upalpha-\upbeta)}{\cosh \upalpha  + \cosh \upbeta } \tilde r,\\
    Q_L = \frac{1}{G_N} \sinh \upalpha \cosh \upbeta  \cdot \frac{D-3}{D-2} \tilde M',  & \quad
    Q_R = \frac{1}{G_N} \cosh \upalpha \sinh \upbeta  \cdot \frac{D-3}{D-2}  \tilde M',\\
    S_\nu = \frac{2\pi R}{G_N} \frac{D-3}{D-2} \tilde M', &\quad M = \frac{1}{G_N}
    \frac{1+(D-3)\cosh \upalpha \cosh \upbeta }{D-2}\tilde M',
\end{split}
\end{equation}
where we used $e^{2\phi_D} = G_N$ in string units.
As expected, the dependence on $n$ disappeared, and the only $\nu$ dependence is through the seed's equation of state for $\tilde M'$ and $\tilde S_\nu'$. 
At fixed dilaton value $\phi_D$, the charged solution is parametrized by the seed's solution through $\tilde M'$, and the transformation parameters $\upalpha$ and $\upbeta$. After substituting the neutral black hole equation of state, one can directly check that \eqref{eq:map_vars_bh} reproduces the equation of state of the 2-charge black hole solution \cite{Sen:1994eb,Horowitz:1995tm}.\footnote{For the leading $\alpha'$ correction see \cite{Massai:2023cis}. It would be interesting to use the general $\nu$ solution-generating method \eqref{eq:map_vars} to find the $\alpha'$ corrections of rotating charged black holes.}

Before discussing the index, let us consider first black hole contributions to the thermal partition function.
In order to match with the BPS entropy \eqref{eq:S_micro}, Sen \cite{Sen:1994eb} looked for an extremal black hole solution with the same charges. 
Such a charged black hole should saturate the supersymmetric BPS bound $M = |Q_R|$. Starting with \eqref{eq:map_vars_bh}, Sen suggested the following scaled limit \cite{Sen:1994eb}
\begin{equation}\label{eq:Sen_limit_1}
    \upbeta\rightarrow\infty, \quad \tilde M' \rightarrow 0, \quad\text{ with }\quad \frac{\tilde S'}{2\pi \tilde R} \cosh \upbeta = \frac{D-3}{D-2} \tilde M' \cosh \upbeta \equiv M_0.
\end{equation}
Since $\phi_D$ is fixed, by \eqref{eq:phi_D_trans}, the seed's string coupling needs to be arbitrarily weak in the limit.
Taking the limit in \eqref{eq:map_vars_bh}, naively gives a BPS solution with
\begin{equation}\label{eq:Sen_limit_res}
\begin{split}
    M = |Q_R| = \frac{M_0}{G_N} &\cosh \upalpha, \quad Q_L = \frac{M_0}{G_N} \sinh \upalpha,
    \quad\text{ and }\quad S = \pi \tilde R \sqrt{Q_R^2-Q_L^2}.
\end{split}
\end{equation}
As already noted by Sen, for $\tilde M' \sim 1$ (in string units), the solution is no longer weakly curved, and the limit \eqref{eq:Sen_limit_res} is unreliable. If one wrongly tries to take the limit using the GR solution, the zero mass limit corresponds to unbounded seed temperature $\tilde R = 0$, which leads in \eqref{eq:Sen_limit_res} to a vanishing area and entropy $ S = 0$.
Such seemingly vanishing-area solutions are sometimes referred to as ``small black holes''.

Note that \eqref{eq:Sen_limit_res} does give the correct entropy dependence on the charges up to an overall $\tilde R$ factor. 
Since the Sen limit breaks around $\tilde R \sim 1$, one could imagine that $\alpha'$ corrections resolve the limit and give the correct answer for the entropy. Such an expectation was first put forward by \cite{Sen:1995in}. Further studies \cite{Dabholkar:2004yr,Dabholkar:2004dq,Hubeny:2004ji,Sen:2004dp} suggested that a string-corrected solution precisely reproduces the entropy. Others \cite{Cano:2018hut, Ruiperez:2020qda}
contested this view, claiming that higher-derivative corrections keep the extremal solution singular, and thus unreliable. Accordingly, it was recently suggested \cite{Chen:2021dsw} that, just like the uncharged black hole, the non-extremal charged black hole transitions to a charged Horowitz-Polchinski solution at lower masses.\newline

We now return to our main goal: studying charged black hole contributions at small $\nu$.
For fixed charges $Q_L, Q_R$ and temperature $R$, we ask which $\tilde M'$,$\upalpha$ and $\upbeta$ contribute for small $\nu$. 
For simplicity, we will assume $D>4$ (the $D=4$ case was studied in \cite{Chen:2024gmc}). Using \eqref{eq:D_ge_4_nu0}, turns out there are two types of solutions:

One option is to fix $\tilde R \approx R$, and take a slightly transformed version of the uncharged solution $\upalpha, \upbeta \ll 1$.
As we saw above \eqref{eq:D_ge_4_nu0}, the seed's mass $\tilde M' \sim \nu^{4-D} \tilde R^{D-3}$ goes to infinity in the $\nu=0$ limit. To fix the charges, one sets $\upalpha \sim \frac{D-2}{D-3} G_N Q_L/\tilde M'$ and $\upbeta \sim \frac{D-2}{D-3} G_N Q_R/\tilde M$, both small for small $\nu$. However, since $\tilde M'$ diverges, so does the charged saddle's mass $M\rightarrow \infty$. Moreover, since the seed's horizon radius $r_h$ diverges, the charged black hole size also grows indefinitely. Overall, this $\nu=0$ limit decouples from the index partition function in the same way as the neutral black hole.

The more interesting option is to take a large $\upbeta \gg 1$ as we take the limit. Fixing the temperature and the charges, one finds $\cosh \upbeta \sim (G_N^{-1} R^{D-3}/|Q|)^{1/(D-4)} \cdot 1/\nu$ and $\tilde M' \sim (G_N |Q|/R)^\frac{D-3}{D-4} \cdot \nu$, with $|Q| \equiv \sqrt{Q_R^2-Q_L^2}$. Thus, this $\nu=0$ limit implements the Sen limit \eqref{eq:Sen_limit_1}, with fixed $\tilde M' \cosh\upbeta \sim (G_N |Q|)^\frac{1}{D-4}$. 
The difference from extremality is
\begin{equation}\label{eq:bh_bps_diff}
\begin{split}
    \frac{M-|Q_R|}{|Q_R|} \sim e^{-2\upbeta} \sim 
    T_{*}^2 \cdot \left(T/T_{*}\right)^\frac{2(D-3)}{D-4}, \quad 
    \frac{S_\nu}{S_\text{micro}} \sim \tilde R \coth \upbeta \sim \left(T/T_{*}\right)^\frac{1}{D-4},
\end{split}
\end{equation}
with $T=1/(2\pi R)$ the temperature, and 
\begin{equation}
    T_{*} = \frac{1}{G_N |Q| \ \nu^{D-4}}.
\end{equation}
Although the second $\nu=0$ limit is very similar to the thermal Sen limit \eqref{eq:Sen_limit_res}, this time (wrongly) it leads to a non-vanishing area solution, but a vanishing index $S_\text{index} \equiv S + 2\pi i J = 0$ \cite{Chowdhury:2024ngg,Chen:2024gmc}.
However, for similar reasons, the limit goes beyond the GR approximation and leads to curvature singularities at $\nu=0$ \cite{Chowdhury:2024ngg,Chen:2024gmc}.
Along the same line as the extremal limit, \cite{Chowdhury:2024ngg,Chen:2024gmc} argued that $\alpha'$ corrections, this time at the supersymmetric point $\nu=0$, can resolve the curvature singularities and reproduce the microscopic index \eqref{eq:S_micro}. 
In this section, we remain agnostic about small black holes and only study weakly curved black hole solutions. We will return to them after discussing the charged W star.

\begin{figure}
    \centering
    \includegraphics[width=0.5\linewidth]{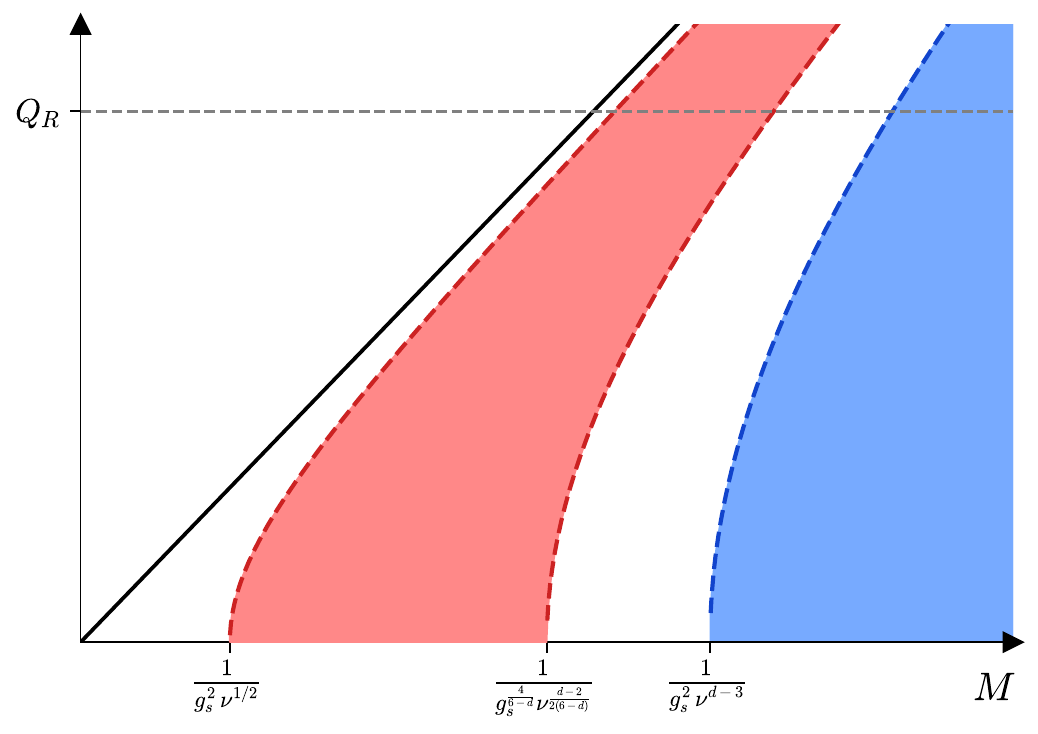}
    \caption{Schematic picture of the different reliable regimes of the mass and the charge $Q_R$ at fixed $\nu$ and $d>4$, in string units. The red and blue regions correspond to the W star and the (weakly curved) black hole, respectively.}
    \label{fig:MQ_space}
\end{figure}

The charged black hole is reliable only for $\tilde R \gg 1$, or $T \gg T_{*}$.
At $T \sim T_{*}$ we find $\cosh \upbeta \sim G_N |Q| \nu^{D-4}$
, and the difference from extremality \eqref{eq:bh_bps_diff} is
\begin{equation}
    \frac{M-|Q_R|}{|Q_R|} \sim \frac{1}{(G_N |Q|\ \nu^{D-4})^2}, \quad \frac{S_\nu}{S_\text{micro}} \sim O(1).
\end{equation}
For a given $\nu$ and large $Q_R \gg 1/\nu^{D-4}$, one can get quite close to the BPS bound. 
However, at fixed charges and small $\nu$, we no longer have $\upbeta \gg 1$ and the solution is very far from the BPS bound.
Put differently, as we decrease $\nu$, the reliable regime of black hole solutions moves further away from BPS and eventually decouples.
In terms of the mass, weakly curved solutions exist only for $\sqrt{M^2-|Q|^2} \gg 1/(G_N \nu^{D-4})$. At fixed charges, this mass regime decouples in the limit, see figure \ref{fig:MQ_space}.

\subsection{Charged W stars}\label{subsec:charged_w_stars}
To leading order in $\tilde R-R_H(\nu) \ll 1$, the uncharged W star satisfies the equation of state $\tilde S_\nu' = 2\pi R_H(\nu) \tilde M'$ \eqref{eq:hag_w_star}. To find the thermodynamics of the charged W star, we substitute the relation in \eqref{eq:map_vars}, which gives 
\begin{equation}\label{eq:w_charged_entropy}
    S_\nu
    =\pi \left(R_H-\frac{n}{R_H}\right)\sqrt{M^2-Q_R^2}
    +\pi \left(R_H+\frac{n }{R_H}\right)\sqrt{M^2-Q_L^2},
\end{equation}
where the charges and the mass are parametrized by
\begin{equation}\label{eq:QMR_W}
\begin{split}
    Q_L = \frac{\tilde M'}{G_N} &\sinh \upalpha  \cosh \upbeta , \quad 
    Q_R = \frac{\tilde M'}{G_N} \cosh \upalpha  \sinh \upbeta ,\quad
    M = \frac{\tilde M'}{G_N} \cosh \upalpha \cosh \upbeta .
\end{split}
\end{equation}
This is a generalization of \cite{Chen:2021dsw} for general $n$ and $R_H$, and should hold for any winding condensate to leading order in $\tilde R-R_H(\nu) \ll 1$. As in \cite{Chen:2021dsw}, \eqref{eq:w_charged_entropy} agrees with the Cardy limit of the free string. 
Finally, to subleading order, the temperature is
\begin{equation}\label{eq:W_charged_temp}
\begin{split}
    R &= \frac{R_H+n/R_H}{2}\cosh \upalpha  +\frac{R_H-n/R_H}{2}\cosh \upbeta \\
    &+ \left(
    \frac{1-n/R_H^2}{2} \cosh \upalpha + \frac{1+n/R_H^2}{2} \cosh \upbeta
    \right)(\tilde R-R_H) + O((\tilde R-R_H)^2).
\end{split}
\end{equation}
Specializing to Heterotic strings and close to $\nu=0$, we have $R_H =1$ and $n=1$. Substituting in \eqref{eq:w_charged_entropy},\eqref{eq:W_charged_temp}, the right-moving part of the entropy vanishes, and we are left with 
\begin{equation}\label{eq:eq_of_state_2}
    S_\nu
    = 2\pi \sqrt{M^2-Q_L^2} + O(\sqrt{\nu}), \quad R\approx \cosh \upalpha + \left(\tilde R - R_H(\nu)+\sqrt{\frac{\nu}{2}}\right) \cosh \upbeta +O(\nu).
\end{equation}
This mirrors the fact that, at the index, the alternating signs cancel any right-moving contributions except the ground state.

At fixed but small $\nu$, the analysis is very similar to the charged Horowitz-Polchinski \cite{Chen:2021dsw,Chen:2024gmc}. 
In the previous section we saw there are two types of black hole solutions. The first was a slightly transformed solution with $R \approx \tilde R$ and small $\upalpha, \upbeta \ll 1$. Since for the W star $\tilde M' \sim \nu^{-1/2}$ diverges in the limit, a similar W star can be found. The description of the transition between the two is a perturbation of the neutral analysis of section \ref{subsec:bh_w_trans}.

The second type of charged black holes involved a large transformation of the seed solution $\upbeta \gg 1$. As in the black hole case, $\nu=0$ imposes a Sen-like limit. From \eqref{eq:eq_of_state_2}, $e^\upbeta \sim R /\nu^{1/2} \gg 1$ and by \eqref{eq:QMR_W} $\tilde M' \sim G_N|Q|/R \cdot \nu^{1/2} \ll 1$. Thus,
for arbitrarily large $\upbeta$, the solution saturate the BPS bound $M\approx|Q_R|$ with entropy 
\begin{equation}
    S_\text{index} \approx 2\pi \sqrt{Q_R^2-Q_L^2} = S_\text{micro}.
\end{equation}
Unlike the large black hole, this naive W star BPS limit agrees with the microscopic entropy \eqref{eq:S_micro}!\footnote{Note that this is not the same computation done in the \cite{Chen:2021dsw}, since here we used the properties of the winding mode close to $\nu=0$.}
To be more precise, the solution approaches the BPS bound as
\begin{equation}\label{eq:w_bps_approach}
    \frac{M-|Q_R|}{|Q_R|} \sim \frac{S_\nu-S_\text{micro}}{S_\text{micro}} \sim e^{-2\upbeta}.
\end{equation}

While the agreement with the microscopic entropy is encouraging, the $\upbeta=\infty$ limit, or the Sen limit, is unreliable just as in the black hole case. 
To understand why, recall that the W star is reliable only within a specific temperature regime.
At lower seed temperatures $\tilde R-R_H(\nu) \sim O(1)$, the uncharged W star arguably transitions into a black hole. By \eqref{eq:eq_of_state_2} we find in this regime $R \sim e^\upbeta$, which corresponds by \eqref{eq:W_low_temp} to a temperature 
\begin{equation}
    T_\text{trans} \sim \frac{1}{g_s^2 |Q|\nu^{1/2}},
\end{equation}
with $g_s^2= e^{2\phi_D}$ the charged solution $D$-dimensional string coupling.
This temperature has a similar, although not exact $\nu$ scaling, structure as the temperature we found above at which the charged black hole turns stringy $T_{*}= \frac{1}{G_N |Q| \ \nu^{D-4}}$ (for $D>4$).
Around $T\sim T_\text{trans}$ the mass and entropy satisfy \eqref{eq:w_bps_approach}
\begin{equation}
    \frac{M-|Q_R|}{|Q_R|} \sim \frac{S_\nu-S_\text{micro}}{S_\text{micro}} \sim T_\text{trans}^2 \sim \frac{1}{g_s^4  \nu |Q|^2}.
\end{equation}

\begin{figure}
    \centering
    \includegraphics[width=0.5\linewidth]{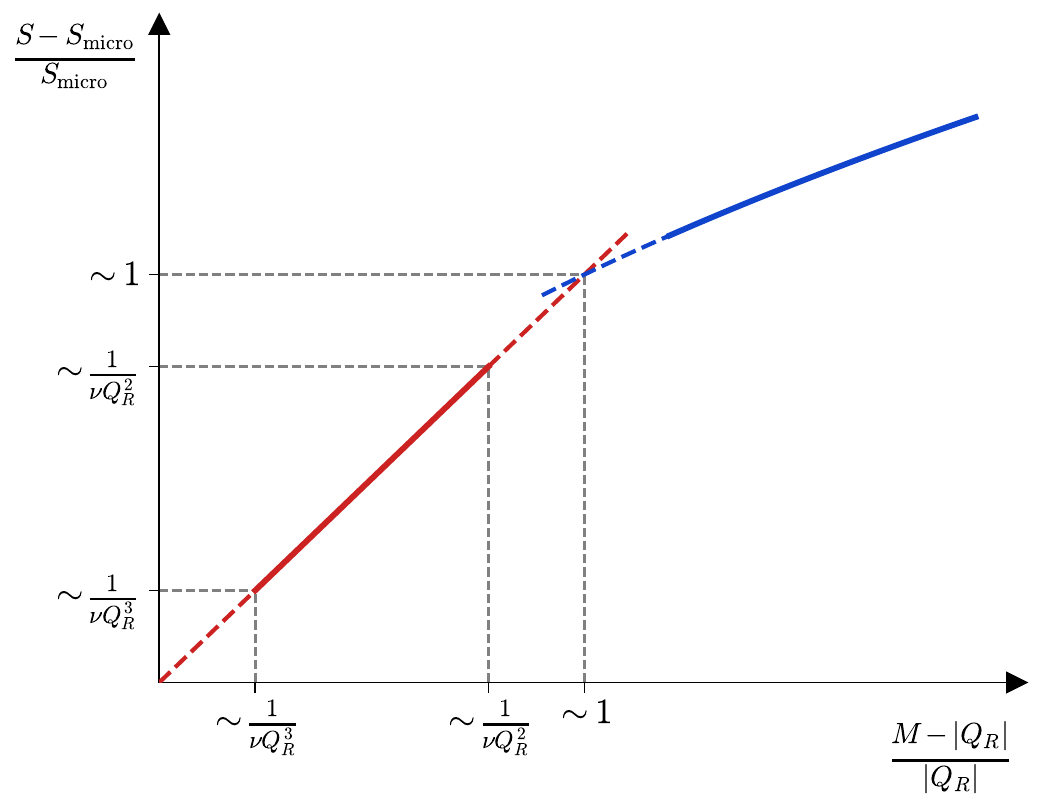}
    \caption{A schematic picture of the transition to BPS, for fixed and large charge $Q_R \gg \sqrt{\nu}$ for $D=4$. The red and blue lines correspond to the W star and the black hole phases, respectively. For small masses, the W star transitions to the free string phase.
    In the $\nu=0$ limit, both the (weakly-curved) black hole and the W star decouple.}
    \label{fig:bps_limit}
\end{figure}

Secondly, at high seed's temperatures, $\tilde R - R_H \sim (\nu^\frac{d-2}{4} \tilde g_s^2)^\frac{2}{6-d}$ \eqref{eq:R_validity_regime}, the solution transitions to free strings.
If $\tilde R-R_H(\nu) \ll \sqrt{\nu}$, we have \eqref{eq:eq_of_state_2} $R \sim \sqrt{\nu} e^\upbeta$ and \eqref{eq:phi_D_trans} $\tilde g_s^2 \sim g_s^2 / R$, which together gives the temperature
\begin{equation}
    T_\text{gas} \sim 
    \frac{1}{g_s^2 |Q|^\frac{6-d}{2}\nu}.
\end{equation}
For consistency, we check
\begin{equation}
    \frac{\tilde R-R_H(\nu)}{\sqrt{\nu}} \sim \frac{1}{|Q| \nu},
\end{equation}
so that we need $|Q| \gg 1/\nu$.
This time, the approach to BPS is
\begin{equation}
    \frac{M-|Q_R|}{|Q_R|} \sim \frac{S_\nu-S_\text{micro}}{S_\text{micro}} \sim \nu \ T_\text{gas}^2 \sim \frac{1}{g_s^4 \nu |Q|^{6-d}}.
\end{equation}
The reliable temperature region depends on the sign of $d-4= D-5$ by (for large $|Q|$)
\begin{equation}
\begin{split}
    T_\text{gas}\ll T\ll T_\text{trans},& \quad d<4\\
    T_\text{trans}\ll T\ll T_\text{gas},& \quad d>4
\end{split}.
\end{equation}
At fixed $\nu$ and large $|Q|$, the solution approaches the BPS bound. Nonetheless, no W star exactly saturates it.
For a schematic plot of the transition at fixed (and large) charges at $d=3$, see figure \ref{fig:bps_limit}.

What happens to this discussion as we take the $\nu=0$ limit? Following the uncharged discussion, the weakly transformed black hole and a W star both decouple in the limit, as they turn arbitrarily heavy and large. 
For the highly transformed solution, at fixed charges, the temperatures $T_\text{trans}$, $T_\text{gas}$, and $T_{*}$ all diverge in the $\nu=0$ limit. Thus, the reliable regimes of temperatures all disappear in the limit. For this reason, the Sen limit is unreliable also for the W star. See figure \ref{fig:MQ_space}.

The $\nu=0$ limit essentially requires us to send the seed mass to zero $\tilde M' \rightarrow 0$. In this limit, however, the (uncharged) W star is arguably transitioning to free strings. As a result, it is natural to suspect that in the $\nu=0$ limit, the charged W star transitions smoothly to a charged free string phase.
Recall that \eqref{eq:w_charged_entropy} already mimics the free string. Moreover, the size of the W star at the transition is $L \sim 1/m \sim (\nu \tilde g_s^2)^{-\frac{1}{6-d}}$, which translates in terms of the charged solution to
\begin{equation}
    L^2 \sim \tilde M /\sqrt{\nu} \sim e^{-2\phi_D} e^\upbeta \tilde M' \sim \sqrt{Q_n Q_w},
\end{equation}
without any $\nu$ dependence. This is the expected size of a generic BPS string with $Q_n$, $Q_w$ charges!
Thus, the transition of the W star at $\nu=0$ to a BPS string phase appears to be smooth. Put differently, while the W star is only approximately BPS with the right entropy behavior, it transitions to an exact BPS string phase in the $\nu=0$ limit. 

As for the black hole, there are two scenarios we can think about. 
If the W star is indeed connected to the black hole saddle, the cleanest option is that only the BPS strings phase survives the limit (possibly with self-gravitating forces as was suggested in \cite{Zigdon:2026thx}, for example).
The other option is that the W star (and similarly the Horowitz-Polchinski solution) is not connected to the black hole at all. In that case, the W star transitions to BPS strings in the limit, while the black hole transitions to the small black hole saddle suggested in \cite{Chowdhury:2024ngg,Chen:2024gmc}.\footnote{
A black hole with $\beta \Omega = 2\pi i$ is a saddle with a contractible $\partial_\tau + 2\pi \partial_\theta$ cycle. As such, one would expect the thermal winding string to have a VEV. However, the analysis in section \ref{sec:het_s1} shows that the winding mass at $\nu=0$ is always non-negative, which indicates no VEV. In other words, there's no asymptotic stringy observable that distinguishes it from the flat space saddle.
}

\section{Type II strings}\label{sec:type_ii}
\begin{figure}[ht]
    \centering
    \includegraphics[width=0.7\linewidth]{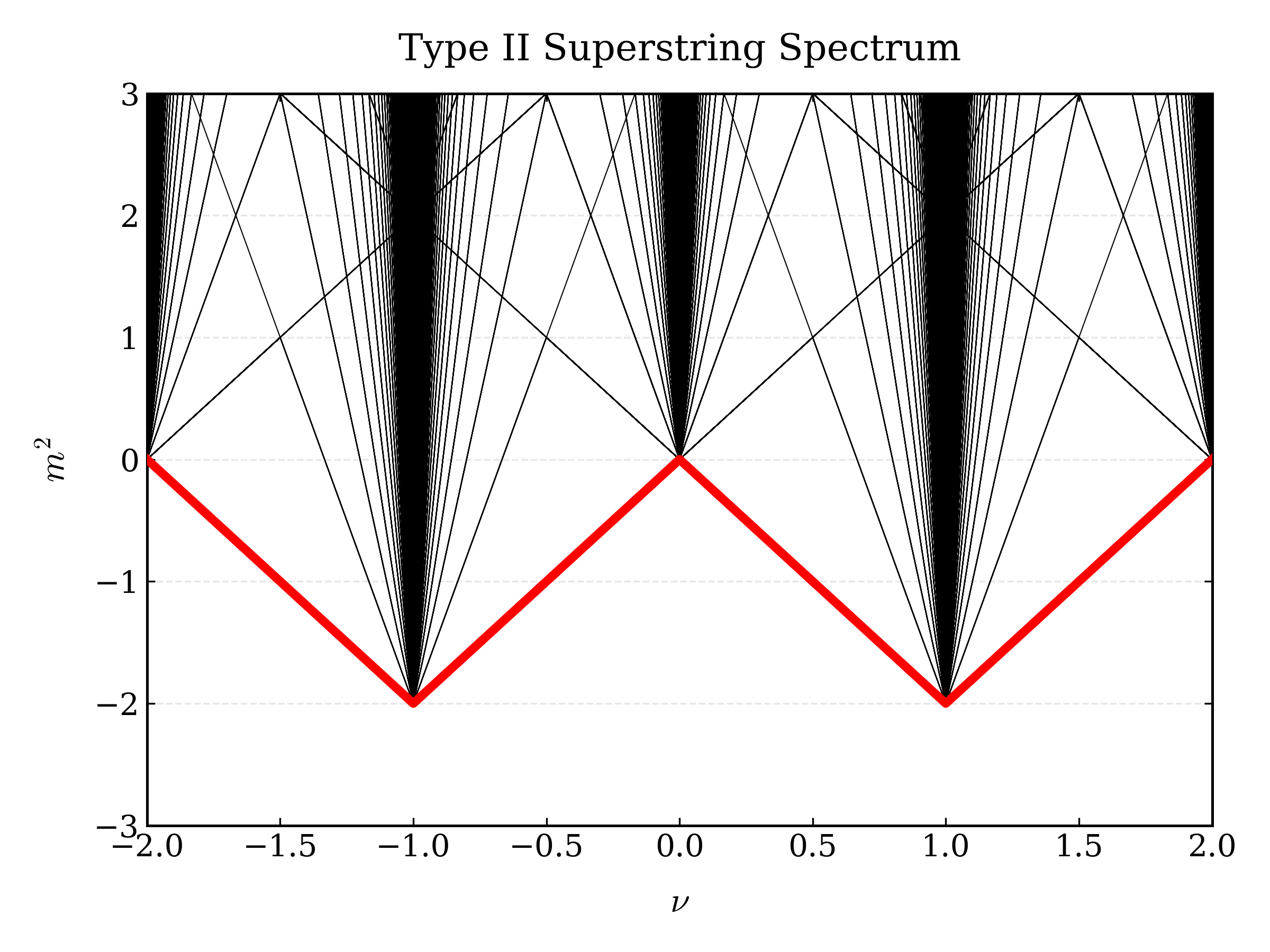}
    \caption{Type II $(d-2)$-dimensional winding $w=1$ spectrum as a function of $\nu \sim \nu+2$. We set $R=0$ at which the classical contribution to the mass vanishes. The spectrum is tachyonic (at small $R$) for any $\nu \ne 1 \text{ mod } 2$.
    It is dense close to $\nu=0$ and $\nu=1$, while for any $\nu$, the winding ground state (in red) is unique.}
    \label{fig:type_ii_spectrum}
\end{figure}

So far, our discussion of the $\nu=0$ limit was limited to Heterotic string theory. In the thermal case, different string theories share a similar near-Hagedorn (Atick-Witten) EFT, due to the universal Hagedorn growth of states. Close to the index, however, the behavior highly depends on the string theory in question. In this section, we comment on the case of type II strings (either IIA or IIB).

\subsection{Worldsheet analysis}\label{subsec:ii_worldsheet}
Let us repeat the analysis of the (pure NS-NS) background \eqref{eq:st_back} for the case of type II strings. 
At thermal momentum $n$ and winding $w$, the type II mass-shell condition in the NS-NS sector\footnote{The other sectors can be shown to never be tachyonic, and don't contribute any non-trivial low energy spectrum.} is
\begin{equation}\label{eq:type_II_mass_shell}
\begin{split}
    m^2 &= (w R+n/R)^2-2(1-w\nu) +4 N \\
    &= (w R-n/R)^2-2(1-w\nu) +4 \tilde N,
\end{split}
\end{equation}
with $w\nu$ in the fundamental domain $0<w \nu<1/2$ (see figure \ref{fig:type_ii_spectrum}). 
The $w=1$ ground state correspond to levels $N=\tilde N= 1/2-\nu$, zero thermal momentum $n=0$ and a mass \cite{Seitz:2025wpc}
\begin{equation}
    m^2 = R^2-2\nu.
\end{equation}
In the thermal case $\nu=1$, the mass vanishes at the known type II Hagedorn temperature $R_H = \sqrt{2}$ \cite{Bowick:1985az}. At the index, the mass is simply 
\begin{equation}\label{eq:ii_m2_index}
    m^2_{\nu=0} = R^2.
\end{equation}
This is the classical worldsheet energy of a winding string, with no quantum corrections due to the worldsheet supersymmetry. As in the Heterotic case, this mass is always non-negative due to the target space supersymmetry.
This time the mass doesn't vanish at the self-dual radius, but at $R=0$! 
Thus, the near-Hagedorn region, in which the winding mode is light, now involves a very small $R \ll 1$.

An important implication of that, compared to the Heterotic case, is that in this regime not only the $w=1$ sector is light, but arbitrarily many winding sectors.
For small $\nu \ll 1$ and any $w$, the light levels are 
\begin{equation}
    N = \frac{1}{2} + w\nu (k-1), \quad \tilde N = \frac{1}{2} + w\nu(\tilde k-1),
\end{equation}
with $k,\tilde k \ge 0$, which correspond to masses and thermal momentum \eqref{eq:type_II_mass_shell}
\begin{equation}\label{eq:spectrum_ii}
    m^2_{k,\tilde k} = w^2 R^2+2w \nu(k+\tilde k-1)+\frac{\nu^2}{R^2}(k-\tilde k)^2, \quad n = (\tilde k - k)\cdot\nu.
\end{equation}
The $w$-sector ground state $k=\tilde k=0$ mass is
\begin{equation}\label{eq:ii_w_m2}
    m^2_w = w^2 R^2 -2\nu w.
\end{equation}
The $w$ sector ground state turns tachyonic for $R^2 = \frac{2\nu}{w}$, and remain light at $R^2 \sim 2\nu$ for $w \ll 1/\sqrt{\nu}$. 

\subsection{The EFT and the string star}\label{subsec:ii_eft_etc}
The following analysis overlaps \cite{David:2001vm}, and we will mostly follow their conventions. As in the Heterotic case, we will assume a compactification to $d$ spatial dimensions.\\

One could hope to dimensionally reduce the thermal circle and find a $d$-dimensional EFT similar to Atick-Witten's or the $SU(2)$ theory we found for Heterotic strings. The result of the last section shows that such a $d$-dimensional theory should have an order $\sim 1/\sqrt{\nu}$ light fields!
The fact that there are arbitrarily many light fields for small $\nu$ means that we can't hope to write down such an EFT. There is, however, a $D=d+1$ dimensional EFT. After all, in the T-dual picture, the different winding sectors correspond to simple thermal momentum sectors. Instead of a small $R\ll 1$, we now have a very large thermal circle $R'=1/R \gg 1$. If we denote by
$p = w/R' = w R$ the momentum in the T-dual picture, the mass \eqref{eq:ii_w_m2} take the form
\begin{equation}
    m^2(p) = p^2 -2 q |p|,
\end{equation}
with $q\equiv\nu R' = \nu/R$. 

For a large thermal circle $R' \gg 1$, the low-energy EFT is known, and it is nothing but the ($D$-dimensionally reduced) Euclidean supergravity action of type II. Before T-duality, our background is the magnetic Melvin background \eqref{eq:melvin_back}, upon compactifying to $R^d$, which in $D$-dimensions is
\begin{equation}
    ds^2= R^2 d\tau^2 + d\rho^2 +\rho^2 (d\theta +\nu d\tau)^2 + dX^2, \quad B=0, \quad \Phi = \Phi_0,
\end{equation}
with $\tau \sim \tau+2\pi$, $dX^2$ stands for $R^{d-2}$, and $\Phi_0$ the $D$-dimensional dilaton. Applying the Buscher rules along $\tau$ gives the T-dual background ($\tau' \sim \tau' + 2\pi$)
\begin{equation}
\begin{split}
    ds^{2\prime} &= \frac{1}{R^2+\nu^2 \rho^2}(d\tau')^2 + d\rho^2 + \frac{R^2\rho^2}{R^2 + \nu^2 \rho^2}d\theta^2 + dX^2,\\
    B' &=\frac{\nu \rho^2}{R^2 + \nu^2 \rho^2} d\tau'\wedge d\theta,\\
    \Phi' &= \Phi_0 -\frac{1}{2} \log \left(R^2+\nu^2 \rho^2\right).
\end{split}
\end{equation}
We can further define $t= R' \tau'=\tau'/R$ ($t' \sim t' + R'$), in which
\begin{equation}\label{eq:q}
\begin{split}
    ds^{2\prime} &= \frac{1}{1+q^2 \rho^2}dt^2 + d\rho^2 + \frac{\rho^2}{1 + q^2 \rho^2}d\theta^2 + dX^2,\\
    B' &=\frac{q \rho^2}{1 + q^2 \rho^2} dt\wedge d\theta,\\
    \Phi' &= \Phi_0 -\frac{1}{2} \log \left(R^2(1+q^2 \rho^2)\right).
\end{split}
\end{equation}
The background is clearly weakly coupled (for small $\Phi_0$), and weakly curved for $q \ll 1$.
Since we know our light winding modes take place in terms of light momentum modes along the circle, one should be able to identify these modes in terms of the momentum modes of the $D$-dimensional NS-NS sector. This was done in \cite{David:2001vm}, in which the authors identify these modes to be
\begin{equation}\label{eq:tachyon}
\begin{split}
    \delta ds^{2'} &= e^{i p t - |p| q \rho^2/2}
    \left(
    \left(d\rho - \frac{i q \rho}{1+q^2 \rho^2} dt\right)^2 + 
    \left(\frac{\rho}{1+q^2 \rho^2} d\theta\right)^2
    \right),\\
    \delta B' &= i e^{i p t - |p| q \rho^2/2}
    \left(d\rho-\frac{i q \rho}{1+q^2 \rho^2} dt\right) \wedge \frac{\rho}{1+q^2 \rho^2} d\theta,\\
    \delta \Phi' &= e^{i p t - |p| q \rho^2/2} \frac{1}{2(1+q^2 \rho^2)}.
\end{split}
\end{equation}
Similar to the Heterotic case, this mode is a Gaussian around $\rho=0$ only with a scale $\rho^2 \sim |p| q$. In the $q=0$ limit, this mode is a superposition of simple planewaves $e^{i p t}$, each with a mass $m^2 = p^2$.

After we identify the tachyonic modes within the EFT, we would like to ask for solutions analog to the W stars. That is, normalizable solutions in which these light momentum modes condense. While this question seems like a complicated one in terms of the full $D$-dimensional theory, we can take a simplifying limit. For a fixed but small $\nu$, the lowest winding mode $w=1$ has a vanishing mass at $R_H^2(\nu) = 2\nu$, and a gap of order $\Delta m^2 \sim \nu$ \eqref{eq:spectrum_ii}. Following section \ref{subsec:lower_E_EFT}, we take the limit where the lowest winding mode is much lighter compared to the gap $m^2 \ll \nu \ll 1$, namely
\begin{equation}
    R-R_H(\nu) \ll \sqrt{\nu} \quad\text{ or }\quad 1/R_H(\nu)-R' \ll  1/\sqrt{\nu}.
\end{equation}
In terms of the EFT, this limit allows us to consider only a single mode with thermal momentum, the $p=1/R'$ case of \eqref{eq:tachyon} which we can call $\chi$. Otherwise, the remaining modes are massless and purely spatial. Upon integrating all the ($d$-dimensional) long-range forces that couple to $\chi$, we will end up with a $(d-2)$-dimensional theory of the type \eqref{eq:eft_w_2} which includes only $\chi$ and its $(d-2)$-dimensional momentum modes. Further analysis is needed in order to determine the exact form of this effective action. 
Yet, the functional form, including the $\nu$ dependence in the coupling, is expected to be the same as in \eqref{eq:eft_w_2}. The analysis of section \ref{sec:W_star_prop} then follows along, and shows that such string stars exist for small $\nu$ also in type II. Moreover, since the neutral black hole solution is independent of the string theory in question, the black hole/ string transition analysis above also holds for the type II solution in the limit that we took.

\subsection{Charged solutions and the BPS limit}\label{subsec:ii_charged}
Finally, we now turn to discuss the charged version of our solutions using the solution-generating technique. Using \eqref{eq:w_charged_entropy} for type II, the entropy to leading order in $\tilde R-R_H \ll 1$ is ($n=0$)
\begin{equation}\label{eq:w_charged_entropy_ii}
    S_\nu
    =\pi \sqrt{2\nu} \left(\sqrt{M^2-Q_R^2}+\sqrt{M^2-Q_L^2}\right).
\end{equation}
At $\nu=1$ (the thermal ensemble) and for BPS states, $M=|Q_R|$ or $M=|Q_L|$, gives $S \approx 2\pi \sqrt{2|Q_n Q_w|}$ in agreement with the free string microscopic index \cite{Chen:2021dsw}. However,  setting $\nu=0$ seems to give a paradoxical $S_\text{index} =0$! Recall that for Heterotic strings, this calculation gave the exact microscopic answer as well, which helped us to argue why the W star transitions to the BPS free string phase.  What is going on?

Let us go back to the free strings, and write the fixed $\nu$ right-moving IIB worldsheet partition function \cite{Seitz:2025wpc}  ($\tau = \tau_1+i \tau_2$)
\begin{equation}
\begin{split}
    Z_\text{R, IIB}(\tau,\nu) &=
    \text{Tr}_\text{R}\left(\exp(2\pi i (\tau_1 P + \tau_2 H + \nu J))\right)\\
    &\sim \frac{1}{\eta(\tau)^9}\frac{\vartheta_{11}(\nu/2,\tau)^4}{\vartheta_{11}(1-\nu,\tau)}\\
    &\sim \frac{e^{\frac{\nu}{B}}}{1-e^{-\frac{2\nu}{B}}+e^{-\frac{2}{B}}},
\end{split}
\end{equation}
where in the third line we substituted $\tau = i \pi B$ and expanded in $B \ll 1$. At fixed $\nu \ne 0$ the partition function diverges as $\exp(\nu/B)$, in line with \eqref{eq:w_charged_entropy_ii}, but in the strict $\nu=0$ it grows as $\exp(2/B)$.
This shows that the $\nu=0$ and the large mass $M,Q \gg 1$ limits don't commute.
$1/2$-BPS are string states with a single (for Heterotic, left) chirality.
At fixed $\nu$, a generic type II single-string high-energy state includes oscillations from both chiralities and entropy $S \sim \pi \sqrt{2\nu} M$, in agreement with \eqref{eq:w_charged_entropy_ii}. Such states are very far from BPS. This is different from Heterotic, where left-moving oscillations become much more likely than right-moving for small $\nu\ll 1$.
But for $1 \ll M \ll 1/\nu$, at smaller and smaller $\nu$, the entropy is now $S \sim \pi\sqrt{2} M$, and a generic state is approximately BPS.
As a function of $M$, there's a crossover for the behavior of a generic single-string state from Hagedorn to BPS.

Going back to our original question, it seems plausible that for sufficiently large $M$, the string star solution transitions to a free string phase. This transition could be smooth for the same reasons we laid out above for Heterotic strings. At lower masses $M \ll 1/\nu$, this phase approaches the BPS entropy. As in the Heterotic case, in the strict $\nu=0$, the string star decouples, and we end up only with a BPS string phase.

It is noteworthy, however, that we found some sort of obstruction when comparing the ``index'' of the type II string star to the free string phase. An obstruction that didn't occur for Heterotic strings. In \cite{Chen:2021dsw}, a sharp disagreement was put forward for type II, and not for Heterotic, between the worldsheet indices of the free string saddle and the black hole. It would be interesting to understand if there is a relation between these findings.

\section{Bosonic strings}\label{sec:bosonic}
\begin{figure}[ht]
    \centering

    \begin{subfigure}{0.48\textwidth}
        \centering
        \includegraphics[width=\linewidth]{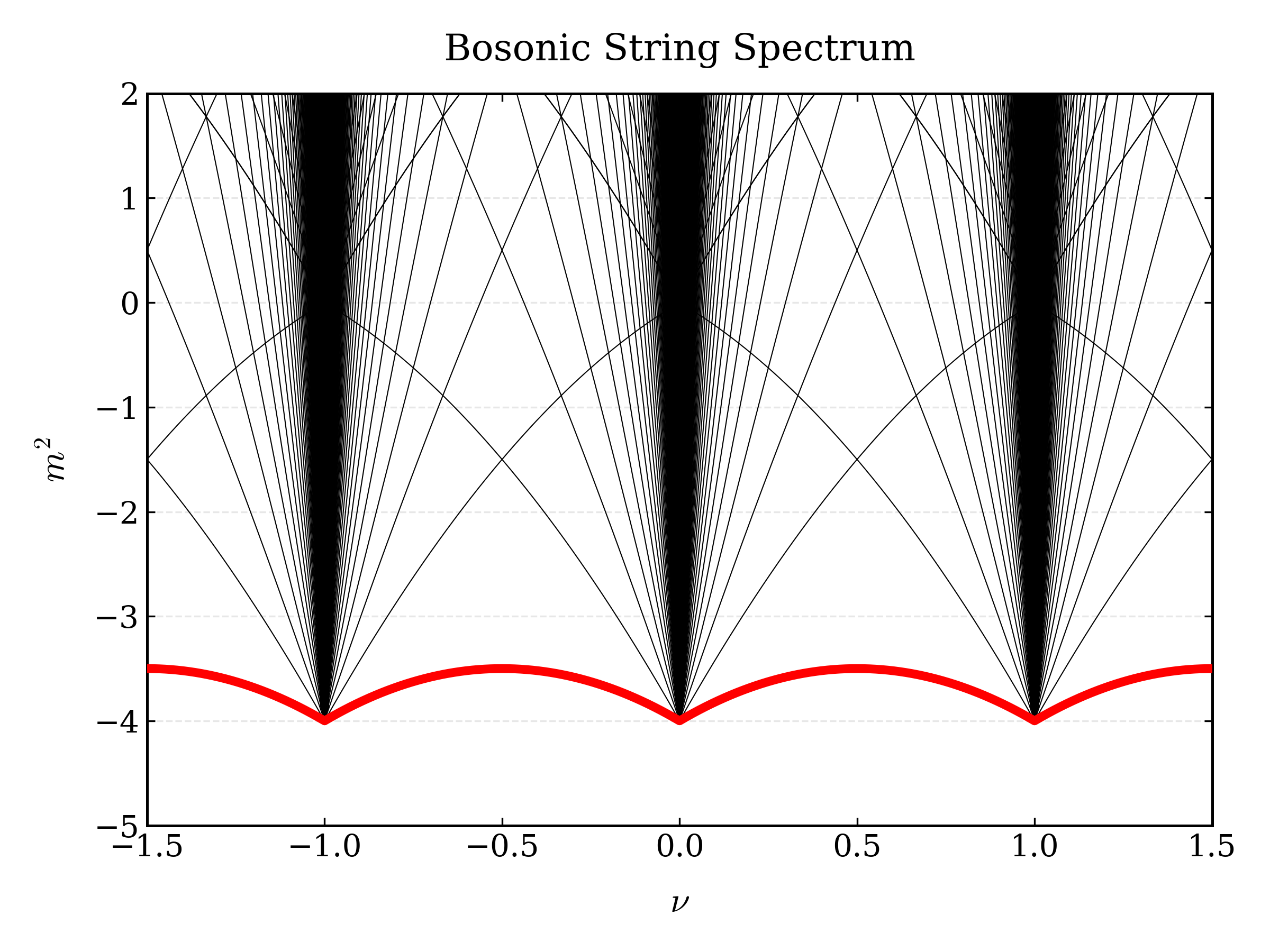}
        \caption{}
        \label{fig:bosonic_spectrum}
    \end{subfigure}
        \hfill
    \begin{subfigure}{0.48\textwidth}
        \centering
        \includegraphics[width=\linewidth]{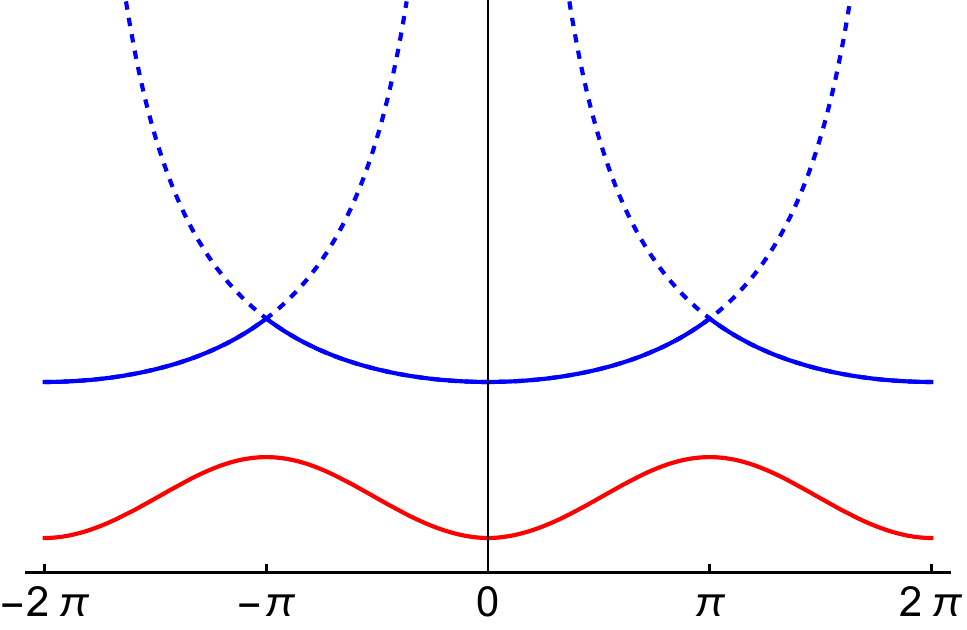}
        \caption{}
        \label{fig:bosonic_free_energy}
    \end{subfigure}
    \caption{\textbf{Left:} Bosonic $(d-2)$-dimensional winding $w=1$ spectrum as a function of $\nu \sim \nu+1$. We set $R=0$ at which the classical contribution to the mass is zero. The spectrum is tachyonic for all $\nu$, and dense close to $\nu=0 \text{ mod } 1$. For any $\nu$, the winding ground state (in red) is unique.\\
    \textbf{Right:} The characteristic free energy of rotating black holes (blue) and the rotating string star (red) as a function of $\omega = -i\beta \Omega \sim \omega+2\pi$, for Bosonic string theory. The black hole free energy was calculated by taking the minimal free energy black hole solution at a given $\omega$, which creates a derivative discontinuity at $\omega = \pm \pi$. The string star free energy (within its regime of validity) is expected to be a continuous function of $\omega$.}
    \label{fig:comparison}
\end{figure}

For the following discussion, we denote $\omega\equiv-i\beta\Omega=2\pi (1-\nu)$.
Bosonic string theory is not supersymmetric, and thus $\omega=2\pi$, or $\nu=0$, doesn't correspond to an index anymore. It is also a bosonic theory, which means that the partition function has a shorter periodicity $\omega \sim \omega+2\pi$. This periodicity is reflected in the Euclidean spectrum of the background, which also goes to itself under  $\omega \sim \omega+2\pi$. Moreover, see figure \ref{fig:bosonic_spectrum}, the winding $w=1$ ground state is unique for any $\omega$, and experiences no spectral flow as we continuously change $\omega \mapsto \omega+2\pi$. By working in the limit $m^2 \ll \Delta m^2$ in which only the winding ground state condenses,
the string star saddle is also unique and goes to itself as we deform $\omega\mapsto \omega+2\pi$.\footnote{The Atick-Witten description is reliable around $\omega \ll 1$ \cite{Seitz:2025wpc}, and as a result also around, say, $|\omega-2\pi|\ll 1$. In order to show that a reliable solution exists, for $m^2 \ll \Delta m^2$, also around $\omega \sim \pi$, a worldsheet calculation should be made to find the couplings in the type II analog of \eqref{eq:eft_w_2}. However, by extrapolating from $\omega\ll 1$, it is reasonable to expect no surprises at $\omega \sim \pi$.} 
One could say that the string star knows that the string spectrum is bosonic.

Let us compare this with the black hole. 
Section \ref{subsec:bh_w_trans} discusses black hole solutions with $-2\pi<\omega<2\pi$. Those solutions do not go to themselves under continuous $\omega \mapsto \omega+2\pi$. After all, the (singular) $\omega=2\pi$ limit is very different from the non-rotating $\omega=0$ black hole. Only after considering all the black hole contributions to the partition functions is the periodicity of the partition function restored. For example, at $\omega=\pi$ both the $\omega=\pm \pi$ black holes contribute. Defining $I_\text{BH}(\omega)$ as the dominant black hole action, gives a derivative discontinuity at $\omega=\pi$ (mod $2\pi$).\footnote{We consider $I_\text{BH}(\omega)$ as the dominant black hole action, and not of a single black hole branch (which would be continuous at $\omega=\pi$). This way, we consider all the sub-dominant contributions as non-perturbative corrections to the same ``black hole phase'' of the system.} By contrast, the string star on-shell action $I_\text{string star}(\omega)$ is continuous at $\omega=\pi \text{ mod }2\pi$ by the aforementioned argument. See figure \ref{fig:bosonic_free_energy} for an illustration.

Let us assume that in the controlled, slowly rotating, regime $\omega \ll 1$, the black hole and the string star saddles are continuously related as a function of the temperature. The unavoidable outcome appears to be that the discontinuity at $\omega=\pi$ should develop at an intermediate temperature. This is confusing, since we usually think about such discontinuities as resolved by non-perturbative $G_N$ effects (as in the Hawking-Page transition, for example).  Since our entire discussion is at tree level, it seems inevitable that stringy effects resolve the discontinuity at sufficiently high temperatures. It is not clear to us if this tension contradicts the black hole/string transition itself (at least for the Bosonic string) or not.

In AdS$_3$, a similar sum over black hole solutions exists in terms of $SL(2,\mathbb{Z})$ mappings\footnote{For Fermionic theories we consider images under the $\Gamma_\theta$ subgroup of $SL(2,\mathbb{Z})$.}, which results in similar discontinuities as a function of the asymptotic modular parameter $\tau = \frac{\omega}{2\pi}+i R$. In \cite{Urbach:2023npi,Seitz:2025wpc} it was suggested that the black holes are continuously related to string stars, while the latter appear continuous for the same reasons laid above. There as well, for the transition to make sense, these discontinuities need to be resolved (at some intermediate $\tau$) by stringy effects.

\section{Discussion and open questions} \label{sec:discussion}
We considered aspects of the black hole/string transition close to the supersymmetric index, by turning on an imaginary angular velocity $\beta \Omega = 2\pi i (1-\nu)$, with $|\nu| \ll 1$. In contrast to the thermal case, the analysis depends heavily on the string theory in question.

For Heterotic strings, we found that instead of the Atick-Witten theory, the effective description is a Higgsed $SU(2)_L\times U(1)_R$ SUGRA, in which the winding mode is the W boson. The Hagedorn instability is manifest (for $\nu \ne 0$) via a version of the Ambjørn--Olesen phenomenon. The string star in this regime is a bubble of W boson condensate, with properties similar to the Horowitz-Polchinski solution. In the $\nu=0$ limit, the string star decouples from the spectrum, just like the weakly-curved black hole.
For type II string theory, an arbitrary number of winding strings are light together for $|\nu| \ll 1$, leading to a new T-dual dimension. The effective $D$-dimensional theory admits a somewhat complicated tachyonic mode for $\nu \ne 0$ \cite{David:2001vm}. Nonetheless, a string star solution arguably exists in an appropriate low-energy limit, leading to similar results.

String star solutions with NS-NS charges can be found by generalizing the method put forward in \cite{Chen:2021dsw}, for both Heterotic and type II string theories. 
The same picture suggested in \cite{Chen:2021dsw} seems to hold for $0<\nu\ll 1$, in which the string star interpolates between the free string and the black hole phase. Following the uncharged case, the charged string star decouples in the $\nu=0$ limit. This suggests that both the black hole and the string star phases decouple, leaving only a BPS string phase. However, this picture is in tension with the small black hole saddle suggested recently for the index \cite{Chowdhury:2024ngg,Chen:2024gmc}.

For Heterotic strings, the transition to the BPS string phase appears to be smooth. Specifically, for large charges, the charged string star entropy approaches the microscopic BPS entropy. For type II, the transition could be smooth, but requires some crossover in the free string phase from a generic highly-excited string and a BPS string. It would be interesting to find a possible connection to the worldsheet index arguments of \cite{Chen:2021dsw}.
We leave these questions for future work.

Using the new effective description for the winding mode near the index, several extensions of this work come to mind. First, we tuned the winding mass to satisfy $m^2 \ll \nu$ for convenience, which allows us to consider only the ground state. It would be interesting to study the full solution at intermediate values of the temperatures $m^2 \sim \nu \ll 1$. Similarly, the Wilson coefficients of the $(d-2)$-dimensional theory can in principle be found for any $\nu$ through a worldsheet computation. This could help to gather evidence that the self-force of the winding mode remains attractive and supports the picture we suggested.
Thirdly, one can consider several angular velocities at the same time, which will lead in the $m^2 \ll \nu$ limit to an even-lower-dimensional theory. For example in $D=5$ we can turn on $\pi i (1+\nu)(J_{12}+J_{34})$, which would result in a zero-dimensional theory for $\chi$.
It could be interesting to check that the effective force in this case is also attractive, and to study properties of the W star. It could also be interesting to turn on thermal Wilson lines for the Heterotic $SO(32)$ gauge fields and study how the solution changes as a result, especially close to new gauge-enhancement points.

Finally, for Bosonic string theory, we spelled out an important qualitative difference between the black hole and the string star phases. Namely, the black hole has a first-order phase transition at $\omega = -i \beta \Omega=\pi$, while the string star has none. Both type II and Heterotic avoid this issue: since both are Fermionic theories, $\omega \sim \omega +4\pi$, the phase transition could occur at $\omega=2\pi$, not $\pi$, and second, both the black hole and, due to supersymmetry, the string star decouple at $\omega = 2\pi$.
A different version of this issue arises for fermionic but non-supersymmetric string theories. While the black hole free energy still diverges at $\omega=2\pi$, we no longer expect the string star to diverge (the partition function still admits a Hagedorn growth). 

\section*{Acknowledgements}
I would like to thank Ofer Aharony, Andreas Blommaert, Yimming Chen, Juan Maldacena, Ohad Mamroud, Upamanyu Moitra, Carmen N\'u\~nez, Josef Seitz, Nikita Sopenko, Antony Speranza, Gustavo Turiaci, and Edward Witten for useful discussions. I also thank Ofer Aharony and Juan Maldacena for their comments on the draft.
This work was funded by the Robert Oppenheimer Endowed Fund and the Fund for Natural Sciences.

\appendix

\section{\titlemath{$D=9$} \titlemath{$\mathcal{N}=1$} gauged supergravity}\label{app:sugra}
The bosonic sector of $D=9$ dimensional theory of 16 supercharges (half-maximal) gauged supergravity is given by \cite{Gates:1984kr,Sezgin:2023hkc} (see also \cite{Schon:2006kz,Geissbuhler:2011mx,Aldazabal:2015yna, Fraiman:2018ebo})
\begin{equation}\label{eq:gauged_sugra}
\begin{split}
    I_E &= \int d^9x e^{-2\phi} \sqrt{g} \left(-\mathcal{R}
    +\frac{1}{12}H^2
    -4\partial_i \phi \partial^i \phi
    -\frac{1}{8} D_i \mathcal{M}_{MN}D^i \mathcal{M}^{MN}
    + \frac{1}{8}\mathcal{M}_{MN} F^M_{ij} F^{N, ij}
    + V,
    \right)
\end{split}
\end{equation}
where $\mathcal{M}\in O(k+1,1;\mathbb{R})/O(k+1)$ ($M,N = 1,...,k+2$), 
\begin{equation}\label{eq:cov_d}
    D_i \mathcal{M}_{MN} = \partial_i \mathcal{M}_{MN} + f_{MP}^Q A^P_i \mathcal{M}_{QN} + f_{NP}^Q A^P_i \mathcal{M}_{MQ},
\end{equation}
and the potential
\begin{equation}\label{eq:V_pot}
    V = \frac{1}{12}f_{MP}^R f_{NQ}^S \mathcal{M}^{MN}\mathcal{M}^{PQ} \mathcal{M}_{RS} +
    \frac{1}{12}f_{MP}^Q f_{NQ}^P \mathcal{M}^{MN}
    +\frac{1}{6} f_{MN}^P f^{MN}_P,
\end{equation}
and indices are raised with the $O(k+1,1;\mathbb{R})$ metric. $f_{MN}^P$ are the structure constants of the gauge group, itself a subgroup of $O(k+1,1;\mathbb{R})$.
In the following, we will ignore the $SO(32)$ vector multiplets since they exactly decouple from the computation.

Let us first consider the trivial case $k=0$, which accounts for the Heterotic theory on a circle 
. The gauge group is $U(1)_L\times U(1)_R \subset O(1,1;\mathbb{R})$ with trivial $f_{MN}^P=0$.
The matrix $\mathcal{M} \in O(1,1;\mathbb{R})$ and gauge field $A^M_i \in U(1)_L\times U(1)_R$ can be parametrize as
\begin{equation}\label{eq:u1_theroy}
    \mathcal{M}_{MN} = \begin{bmatrix}
        \cosh(\sigma) & \sinh(\sigma)\\
        \sinh(\sigma) & \cosh(\sigma)
    \end{bmatrix}, \quad A^M_i = \begin{bmatrix}
        A_i\\
        \bar A_i
    \end{bmatrix}.
\end{equation}
The indices $M,N=1,2$ are raised by the $O(1,1)$ metric
\begin{equation}
    L_{MN} = \begin{bmatrix}
        1 & 0\\
        0 & -1
    \end{bmatrix}.
\end{equation}
A simple calculation shows that
\begin{equation}
\begin{split}
    -D_i \mathcal{M}_{MN}D^i \mathcal{M}^{MN} & = 
    -2\partial_i (\cosh(\sigma))\partial^i(\cosh(\sigma))
    +2\partial_i (\sinh(\sigma))\partial^i(\sinh(\sigma))\\
    &= 2 \partial_i \sigma \partial^i \sigma.
\end{split}
\end{equation}
and \eqref{eq:gauged_sugra} reduces to the supergravity action \eqref{eq:KK_2}.
\newline

To study the enhanced gauge symmetry, we instead set $k=2$, which corresponds to the two new vector multiplets. As a result, instead of the group $O(1,1;\mathbb{R})$, we now have $O(3,1;\mathbb{R})/O(3)$ \cite{Fraiman:2018ebo} with the $O(3,1;\mathbb{R})$ metric
\begin{equation}
    L_{MN} = \begin{bmatrix}
        1_3 & \ \\
        \  & -1
    \end{bmatrix}.
\end{equation}
The most general $\mathcal{M}_{MN} \in O(3,1)/O(3)$ matrix (connected to the identity) is the ``Lorentz transformation'' with rapidity vector $S_\alpha$, $M=1,...,4$ and $\alpha=1,2,3$,
\begin{equation}
    \mathcal{M}_{MN} = 
        \left[
        \begin{array}{c|c}
        \cosh |S| \delta_{\alpha \beta} + (1-\cosh |S|) (\delta_{\alpha \beta}-\hat S_\alpha \hat S_\beta) & \sinh|S| \cdot \hat S_\alpha\\
        \hline
        \sinh|S| \cdot \hat S_\beta & \cosh|S|
        \end{array}
        \right],
\end{equation}
and $\hat S_\alpha \equiv S_\alpha/|S|$.
As our gauge group, we are instructed to take $SU(2)_L\times U(1)_R$ with the only non-trivial structure constants $f^\gamma_{\alpha \beta}=\epsilon_{\alpha \beta}^\gamma$. We parametrize the gauge fields as
\begin{equation}
    A^M_i = \begin{bmatrix}
        A^\alpha_i\\
        \bar A_i
    \end{bmatrix}.
\end{equation}

Plugging in \eqref{eq:gauged_sugra}, the gauge kinetic terms are 
\begin{equation}
\begin{split}
    \frac{1}{8}\mathcal{M}_{MN} F^M_{ij} F^{N, ij} & = 
    \frac{1}{8} \cosh |S| \left(F^2 + \bar F^2\right)
    + \frac{1}{8}(1-\cosh |S|)[F, \hat S]^2
    + \frac{1}{4} \sinh |S| (\hat S \cdot F) \bar F,
\end{split}
\end{equation}
with implicit $SU(2)$ and spacetime contractions.
Using \eqref{eq:cov_d} we find the derivatives
\begin{equation}
\begin{split}
    D_i \mathcal{M}_{44} &= \partial_i (\cosh|S|)
    = \sinh|S| \hat S^\alpha D_i S^\alpha,\\
    D_i \mathcal{M}_{4,\alpha} 
    &=  \partial_i \mathcal{M}_{4,\alpha} + \epsilon_{\alpha \beta \gamma} A^\beta_i \mathcal{M}_{4,\gamma},\\
    &= \frac{\sinh|S|}{|S|} \left(\delta_{\alpha \beta} + \left(|S|\coth|S|-1\right)  \hat S_\alpha \hat S_\beta \right) D_i S^\beta,\\
    D_i \mathcal{M}_{\alpha,\beta} &=  \partial_i M_{\alpha \beta} 
    + \epsilon_{\alpha \gamma \delta} A^\gamma_i \mathcal{M}_{\delta \beta}
    + \epsilon_{\beta \gamma \delta} A^\gamma_i \mathcal{M}_{\alpha \delta},\\
    &=\left(\sinh|S|-2\cosh|S|+2\right) \hat S_\gamma D_i S^\gamma \hat S_\alpha \hat S_\beta + \frac{\cosh|S|-1}{|S|} \left(\hat S_\alpha D_i S_\beta+\hat S_\beta D_i S_\alpha\right).
\end{split}
\end{equation}
Therefore, the kinetic terms are
\begin{equation}
\begin{split}
    -\frac{1}{8} D\mathcal{M}_{MN}D^i\mathcal{M}^{MN} &= \frac{1}{8}\left(2 (\partial \mathcal{M}_{4,\alpha})^2-(\partial \mathcal{M}_{44})^2 - (\partial \mathcal{M}_{\alpha \beta})^2\right)
    \\
    &=  \frac{1}{4}(DS)^2 + \frac{1}{4}\left(\frac{2\cosh|  S|-2}{S^2}-1\right) [DS,\hat S]^2.
\end{split}
\end{equation}
Lastly, plugging inside \eqref{eq:V_pot} the potential term vanishes. This agrees with the exact worldsheet, in which we know that (in the absence of background gauge fields) the $SU(2)$ adjoint scalar is an exact modulus.

Altogether, the $SU(2)_L\times U(1)_R$ gauge SUGRA theory \eqref{eq:gauged_sugra} is
\begin{equation}\label{eq:gauged_sugra_su2}
\begin{split}
    I_E &= \int d^9x e^{-2\phi} \sqrt{g} \left(-\mathcal{R}
    +\frac{1}{12}H^2
    -4(\partial \phi)^2
    \right.\\
    & \quad
    +\frac{1}{8} \cosh |S| \left(F^2 + \bar F^2\right)
    +\frac{1}{4}(DS)^2 
    + \frac{1}{4}\left(\frac{2\cosh|  S|-2}{S^2}-1\right) [DS,\hat S]^2
    \\
    &\quad 
    \left.
    + \frac{1}{8}(1-\cosh |S|)[F, \hat S]^2
    + \frac{1}{4} \sinh |S| (\hat S \cdot F) \bar F
    \right)
    .
\end{split}
\end{equation}

\paragraph{The unitary-gauge constraint}
In unitary gauge \eqref{eq:unitary_gauge} we fix the values of $S^{1,2}$. As a result, the corresponding gauge constraint is simply the Euler-Lagrange equation for $S^{1,2}$. To find To find it, we first expand \eqref{eq:gauged_sugra_su2} to linear order in $S^1,S^2$ and set $S^3=\sigma$:
\begin{equation}
\begin{split}
    \frac{1}{4}(DS)^2\mid_\text{lin}
    &= \frac{1}{2}\left(\sigma A^2 \partial - \sigma A^3 A^1 - \partial \sigma A^2\right) S^1\\
    &\quad + \frac{1}{2}\left(-\sigma A^1 \partial - \sigma A^3 A^2 + \partial \sigma A^1\right) S^2\\
    \frac{1}{4} \left(\frac{2\cosh|S|-2}{S^2} -1\right)[DS,\hat S]^2 \mid_\text{lin} &= \left(\frac{2\cosh\sigma-2}{\sigma^2} -1\right)\cdot \left(\frac{1}{4}(DS)^2\mid_\text{lin}\right)\\
    \frac{1}{8}(1-\cosh|S|)
    ([F,\hat S])^2 \mid_\text{lin} &=
    \frac{\cosh \sigma-1}{4\sigma}\left(S^1 F^1 + S^2 F^2\right) \cdot F^3\\
    \frac{1}{4}\sinh|S| (S\cdot F) \bar F\mid_\text{lin} &=
    \frac{\sinh(\sigma)}{4\sigma}\left(S^1 F^1 + S^2 F^2\right)\cdot \bar F.
\end{split}
\end{equation}
which gives the constraints (defining $\sqrt{G} = e^{-2\phi}\sqrt{g}$)
\begin{equation}
\begin{split}
    -\frac{1}{\sqrt{G}}\partial_i \left(\sqrt{G} (\cosh(\sigma)-1) A^{2,i}\right) & - (\cosh(\sigma)-1) A^{3,i} A^1_i + \frac{F^{1,ij}}{4} \left((\cosh(\sigma)-1)F^3_{ij}+\sinh(\sigma) \bar F_{ij}\right)=0,\\
    \frac{1}{\sqrt{G}}\partial_i \left(\sqrt{G} (\cosh(\sigma)-1) A^{1,i}\right) &
    - (\cosh(\sigma)-1) A^{3,i} A^2_i + \frac{F^{2,ij}}{4} \left((\cosh(\sigma)-1)F^3_{ij}+\sinh(\sigma) \bar F_{ij}\right)=0.
\end{split}
\end{equation}
In terms of $W,A$ and $\bar A$ \eqref{eq:W_def},
\begin{equation}\label{eq:gauge_const}
\begin{split}
    \frac{1}{\sqrt{G}} (\partial_j +i A_j)&\left(\sqrt{G} (2\cosh(\sigma)-2) W^j\right) - \frac{i}{2} \left((\cosh(\sigma)-1)F^{jk}+\sinh(\sigma) \bar F^{jk}\right)(DW)_{jk}=0.
\end{split}
\end{equation}
Around the standard (supersymmetric) background \eqref{eq:S1_back} we find the Lorentz condition $\partial^j W_j = 0$.

\section{The Atick-Witten EFT}
\label{app:thermal_eft}
In this appendix, we review the target space analysis of the Atick-Witten EFT in the presence of angular velocity potential \cite{Seitz:2025wpc}. We will recast the analysis in a standard KK reduction form that will hopefully help the reader compare the situation to the index case. For brevity, we define $\hat \nu=1-\nu$, so that $\hat \nu=0$ stands for the thermal background (and $\hat \nu=1$ for the index).

Writing down the metric in KK form ($x^0 \sim x^0+2\pi$) \eqref{eq:kk_red}
\begin{equation}\label{eq:kk_red_2}
    ds^2 = g_{ij} dx^i dx^j+ e^{\sigma} \left(dx^0 + a_i dx^i\right)^2, \quad b_i = B_{0i},
\end{equation}
the Atick-Witten 9d quadratic theory for $\chi$ reads
\begin{equation}\label{eq:KK_3}
    I_\chi^{(2)} = \frac{1}{8 G_N} \int d^{9} x e^{-2\phi} \sqrt{g} \left(
    |D \chi|^2
    + \left(e^\sigma+n^2 e^{-\sigma}-c\right)|\chi|^2
    \right),
\end{equation}
with $\phi = \Phi-\frac{1}{4}\sigma$ ($\Phi$ being the $10$-dimensional dilaton),
\begin{equation}
    c = \begin{cases}
        4, & \text{Bosonic}\\
        2, & \text{Type II}\\
        3, & \text{Heterotic}
    \end{cases}
    ,
\end{equation}
and $D_j = \partial_j - i w b_j - i n a_j$.
The field $\chi$ has thermal winding $w=1$ and momentum
\begin{equation}
    n = \begin{cases}
        0, & \text{Bosonic, type II}\\
        \frac{1}{2}, & \text{Heterotic}
    \end{cases}.
\end{equation}

The background we consider is the Magnetic Melvin background
\begin{equation}\label{eq:melvin_back_3}
\begin{split}
    ds^2 & = d\rho^2 + \frac{\rho^2}{1+\hat \nu^2 \rho^2/R^2} d\theta^2 + dX^2,\\
    \sigma &= \log \left(R^2 + \hat \nu^2 \rho^2\right), \\
    a &= \frac{\hat \nu \rho^2}{R^2+\hat \nu^2 \rho^2} d\theta, \quad b = 0,\\
    \phi &= -\frac{1}{4} \log \left(R^2 + \hat \nu^2 \rho^2\right).
\end{split}
\end{equation}
Substituting the background in \eqref{eq:KK_3}, together with the ansatz $\chi = e^{-i p x - i l \theta} \hat \chi(\rho)$ gives
\begin{equation}
    I_\chi^{(2)} = \frac{\Omega_{d-1}}{8 G_N} \int \rho d\rho \left(
    |\hat \chi'|^2
    + \left(
    p^2 + R^2 +\frac{(\hat \nu l -n)^2}{R^2} - c
    +
    \frac{l^2}{\rho^2} + \hat \nu^2 \rho^2 \right)|\hat \chi|^2
    \right).
\end{equation}
Diagonalizing the spectrum in terms of the Landau levels basis $w_{k,\tilde k}$, we find $l=k-\tilde k$ and
\begin{equation}
    \lambda_{p,k,\tilde k} = p^2 + R^2 + \frac{(n-\hat \nu (k-\tilde k))^2}{R^2}-c + 2\hat \nu (1+k+\tilde k),
\end{equation}
with $k,\tilde k\ge 0$. This reproduces the low-energy spectrum, up to an $O(\hat \nu^2)$ higher-derivative shift.

\section{The Ambjørn--Olesen phenomenon}\label{app:adjoint_higgs}
In this appendix, we will review the Ambjørn--Olesen phenomenon of W bosons in a constant magnetic field.
We use Euclidean signature on $R^d$ ($d>2$), to make clear contact with the analysis above. The continuation of the results to Lorentzian signature case is trivial.
We will use complex coordinates $z,\bar z = x^1 \pm i x^2$ for the two-dimensional magnetic plane and denote $x^i$, $i=3,...,d$, for the transverse directions.

Upon breaking the $SU(2)$ Lagrangian $\frac{1}{4} \text{tr}(F^2)$ to $U(1)$ by some Higgs VEV, the quadratic W boson has the following action in unitary gauge\footnote{In our conventions, the covariant derivative is $D= d + i A$, which is the opposite sign compared to the particle physics convention. Therefore, the holomorphic and anti-holomorphic variables are flipped.}
\begin{equation}\label{eq:L_mag}
    I = \int d^d x \left(\frac{1}{4}|(d+i A)\wedge W|^2 + \frac{i}{2} F^{ij} W_{i}W^*_{j} + \frac{1}{2}m^2 |W|^2\right),
\end{equation}
with the $U(1)$ magnetic field $B$ background
\begin{equation}
    A = i \frac{B}{4}(z d\bar z - \bar z dz),\quad F = i \frac{B}{2} dz \wedge d\bar z,
\end{equation}
and W is subject to the gauge constraint
\begin{equation}\label{eq:mag_gauge_const}
    (d+i A)\cdot W = 0.
\end{equation}

Transverse polarizations of W have the general form $W = e^{-i p_j x^j} w(z,\bar z) dx^j$, for a fixed $j \ne 1,2$. By the constraint, we also find $p\cdot W = p^j = 0$. Each of these polarizations behaves as a scalar under the magnetic field
\begin{equation}
\begin{split}
    I &= \int d^2 z \left(\frac{1}{2}|(d+i A)w|^2 + \frac{1}{2}(p^2+m^2) |w|^2\right)\\
    &= \int d^2 z \left(
    \left|\left(\partial_z+ B \bar z/4\right)w\right|^2 + 
    \left|\left(\partial_{\bar z}- B z/4\right)w\right|^2
    + \frac{1}{2}(p^2+m^2) |w|^2\right).
\end{split}
\end{equation}

Let us recall how to construct the Landau levels that diagonalize this action.
The following pairs of creation and annihilation operators
\begin{equation}\label{eq:mag_as}
\begin{split}
    a = \frac{B}{4} z + \partial_{\bar z}, & \quad 
    a^\dagger = \frac{B}{4} \bar z -\partial_z, \\
    \tilde a = \frac{B}{4} \bar z + \partial_z, &\quad 
    \tilde a^\dagger = \frac{B}{4} z-\partial_{\bar z},
\end{split} 
\end{equation}
satisfy the commutation relations
\begin{equation}\label{eq:comm_rel_app}
    [a, a^\dagger] = [\tilde a, \tilde a^\dagger] = \frac{B}{2},
\end{equation}
while the rest of the commutators vanish.
In terms of these operators, the quadratic action takes the harmonic oscillator form
\begin{equation}
\begin{split}
    I_2 
    &= \int d^2 z \ \frac{1}{2} w^*\left(p^2+m^2 +2 \tilde a^\dagger \tilde a + 2\tilde a \tilde a^\dagger\right)w\\
    &= \int d^2 z \ \frac{1}{2} w^*\left( 
    p^2+m^2+B
    + 4 \tilde a^\dagger \tilde a 
    \right)w.
\end{split}
\end{equation}
The Landau-levels ground state (assuming $B>0$)
\begin{equation}
    \label{eq:w_gs_wavefunction_B}
    w_{0,0}(z,\bar z) = \sqrt{\frac{B}{2\pi}} \exp(-B |z|^2/4),
\end{equation}
is defined to be annihilated by both $a \ w_{0,0} = \tilde a \ w_{0,0} = 0$. The rest of the Landau levels are constructed using the creation operators
\begin{equation}
    w_{k,\tilde k} = \frac{(a^\dagger)^{k}(\tilde a^\dagger)^{\tilde k}}{\sqrt{k! \ \tilde k!}} w_{0,0},
\end{equation}
with $k,\tilde k\ge 0$. This is a complete basis.
By the commutation relations \eqref{eq:comm_rel_app}, the corresponding eigenvalues are
\begin{equation}
    \lambda_{k,\tilde k} = p^2 + m^2 + B \left(1+ 2\tilde k\right).
\end{equation}
The spectrum of the theory is independent of $k$, which indicates an enormous degeneracy.
\newline

The longitudinal polarizations $W_1,W_2$ have a derivative coupling to the transverse polarizations, so that the most general longitudinal mode is of the type
\begin{equation}
\begin{split}
    W & = e^{-i p_j x^j} \left(w_+ dz + w_- d\bar z -i p_j dx^j \cdot \phi \right),
\end{split}
\end{equation}
with $w^\pm$ and $\phi$ functions of $z,\bar z$, and subject to the gauge constraint \eqref{eq:mag_gauge_const}
\begin{equation}\label{eq:g_c_long}
    \frac{1}{2} p^2 \phi = D_z w_-+D_{\bar z} w_+,
\end{equation}
with $D\equiv d + i A$ on the RHS.

The different terms in the quadratic action 
\begin{align}
    DW_{z \bar z} &= D_z w_- - D_{\bar z} w_+,\\
    DW_{z,j} &= -i p_j \left( D_z \phi-w_+\right),\\
    DW_{\bar z,j} &= -i p_j \left( D_{\bar z} \phi-w_-\right),\\
    \frac{m^2}{2} W^2 &= m^2 \left(|w_+|^2 + |w_-|^2 + \frac{1}{2} p^2 |\phi|^2\right),\\
    \frac{i}{2} F^{ij}W_i W^*_j &= B \left(|w_+|^2 - |w_-|^2\right).\label{eq:zeeman_w}
\end{align}
can be organized in the following form
\begin{equation}
\begin{split}
    I 
    &=\int d^2 z \Bigg\{
    2 |D_z w_- - D_{\bar z} w_+|^2 + (p^2+m^2+B) |w_+|^2 +(p^2+m^2-B) |w_-|^2\\
    & \quad \quad \quad\quad \quad
    + p^2 \left(|D_z \phi|^2 + |D_{\bar z} \phi|^2 +\frac{1}{2}m^2 |\phi|^2
    +\phi^* (D_z w_-+D_{\bar z} w_+) + c.c.\right)
    \Bigg\}.
\end{split}
\end{equation}
Due to the gauge constraint \eqref{eq:g_c_long} the action can be further organized in an explicitly diagonal form
\begin{equation}
\begin{split}
    I = \int d^2 z \Bigg\{
    &4|D_{\bar z} w_+|^2 + (p^2 + m^2+B) |w_+|^2 \\
    +& 4|D_z w_-|^2 +(p^2 + m^2-B) |w_-|^2\\
    +& \frac{1}{2}p^2 \left(2|D_z \phi|^2 + 2|D_{\bar z} \phi|^2 +(p^2+m^2) |\phi|^2 \right)
    \Bigg\},
\end{split}
\end{equation}
or in terms of the operators \eqref{eq:mag_as}
\begin{equation}
\begin{split}
    I = \int d^2 z \Bigg\{
    &w_+^* \left(2\tilde a^\dagger \tilde a + 2\tilde a \tilde a^\dagger + p^2 + m^2+2B\right) w_+ \\
    +& w_-^* \left(2\tilde a^\dagger \tilde a + 2\tilde a \tilde a^\dagger  + p^2 + m^2-2B\right) w_-\\
    +& \frac{1}{2} p^2 \phi^* \left(2\tilde a^\dagger \tilde a + 2\tilde a \tilde a^\dagger +p^2+m^2 \right) \phi
    \Bigg\}.
\end{split}
\end{equation}
The quadratic operator for both $w_\pm$ and $\phi$ is identical to the scalar case, only with a shifted mass term. In the basis $w_{k,\tilde k}$ for $w_\pm$, we find the eigenvalues
\begin{equation}
    \lambda^{(\pm)}_{k,\tilde k} = p^2 + m^2 + B(2\tilde k+1) \pm 2B,
\end{equation}
while $\phi$ can always be found to satisfy the constraint 
\begin{equation}\label{eq:g_c_a}
    \frac{1}{2} p^2 \phi = \tilde a w_- - \tilde a^\dagger w_+,
\end{equation}
which gives the same eigenvalue. The $\pm 2B$ Zeeman splitting is due to the intrinsic spin coupling to the magnetic field \eqref{eq:zeeman_w}.
The lightest W eigenvalue comes from any $\tilde k=0$ mode of $w_-$,
\begin{equation}
    \lambda^{(-)}_{k,0} = p^2 + m^2-B,
\end{equation}
namely $w_- = w_{k,0}$ and $\phi \sim \tilde a w_- = 0$ by the the gauge constraint. The general vacuum solution (with $\tilde a w_- = 0$) is $w_- = F(\bar z) \exp(-B |z|^2/4)$.
If the magnetic field is strong enough $B > m^2$, the W boson is tachyonic and condenses. This is the Ambjørn--Olesen phenomenon \cite{Ambjorn:1988tm,Ambjorn:1988fx,Ambjorn:1989bd}.

\section{Generating charged solutions} \label{app:sol_gen}
In this appendix, we generalize section 5.2 of \cite{Chen:2021dsw} for any $\nu$ and general winding mode's thermal momentum $n$.
We work in the coordinates where
\begin{equation}
    (t,y)\sim(t,y+2\pi)\sim (t+2\pi,y),
\end{equation}
and the seed solution metric is asymptotically $S^1_R \times S^1_y$
\begin{equation}
    ds^2 = \tilde R^2 dt^2+\tilde r^2 dy^2.
\end{equation}
In order to map the winding mode to itself, we look for a transformation $\Omega\in O(2,2;\mathbb{R})$ with
\begin{equation}
    \Omega \cdot 
    \begin{pmatrix}
        1\\0\\n\\0
    \end{pmatrix} 
    = 
    \begin{pmatrix}
        1\\0\\n\\0
    \end{pmatrix}.
\end{equation}
One basis we can choose is
\begin{equation}
    \gamma_1 = \begin{pmatrix}
        0 & 0 & 0 & \frac{1}{\sqrt{n}}\\
        \sqrt{n} &0 & -\frac{1}{\sqrt{n}}& 0\\
        0&0&0& -\sqrt{n}\\
        0&0&0&0
    \end{pmatrix}, \quad
    \gamma_2 = \begin{pmatrix}
        0 & \frac{1}{\sqrt{n}} & 0 & 0\\
        0 &0 & 0& 0\\
        0&-\sqrt{n}&0& 0\\
        \sqrt{n}&0&-\frac{1}{\sqrt{n}}&0
    \end{pmatrix},
    \gamma_3 = \begin{pmatrix}
        0 & 0 & 0 & 0\\
        0 &2 & 0& 0\\
        0&0&0& 0\\
        0&0&0&-2
    \end{pmatrix},
\end{equation}
which satisfy
\begin{equation}
    [\gamma_1,\gamma_2] =  \gamma_3, \quad 
    [\gamma_3,\gamma_1] = 2\gamma_1, \quad 
    [\gamma_3,\gamma_2] = -2\gamma_2.
\end{equation}
We see that for every $n>0$ the algebra is $\text{sl}(2)$. Using $\Omega = \exp\left(\frac{\sqrt{n} u}{1-n uv} \gamma_2\right) \cdot \exp(\sqrt{n} v \gamma_1)$, the transformed asymptotic metric $B$ field
\begin{equation}
    ds^2 = R^2 dt^2 + r^2(dy+a dt)^2, \quad B_{12}=b
\end{equation}
reads
\begin{equation}\label{eq:new_metric}
\begin{split}
    R^2 &= \left(1-n u v\right)^2 \frac{\tilde R^2 \tilde r^2}{(1+\tilde r^2\tilde R^2 v^2)
    (\tilde r^2 + \tilde R^2 u^2)},\\
    r^2 &= \frac{\tilde r^2 + \tilde R^2 u^2}{(1-n u v)^2(1+\tilde r^2\tilde R^2 v^2)},\\
    a &= i \mu_n R = - \frac{(1-nuv)(\tilde R^2 u +n \tilde r^2 v)}{\tilde r^2+ \tilde R^2 u^2},\\
    b &= i \mu_w R = -\frac{n u +\tilde r^2 \tilde R^2 v}{(1-n u v )(1+\tilde r^2\tilde R^2 v^2)}.
\end{split}
\end{equation}
The $D$-dimensional dilaton transformed so that in $d$-dimensions it remains the same, and so
\begin{equation}
    e^{2\phi_D} = e^{2\tilde \phi_D} \frac{R}{\tilde  R}.
\end{equation}
By inverting the relations, the partial derivatives satisfy
\begin{equation}
    \frac{\partial \tilde R}{\partial R} = \frac{\tilde R}{R} \frac{1+n u v}{1-n u v}, \quad 
    \frac{\partial \tilde R}{\partial \mu_n} = - i \frac{r}{\tilde r}\frac{\tilde R^2 u}{1+\frac{u^2 \tilde R^2}{\tilde r^2}}, \quad
    \frac{\partial \tilde R}{\partial \mu_w} = - i \frac{\tilde r}{r}\frac{ \tilde R^2 v }{1+ v^2 \tilde r^2 \tilde R^2}.
\end{equation}

We begin with the equality of the action under the transformation
\begin{equation}\label{eq:Z_eq_2}
    \log Z(R,r,\mu_n,\mu_w) = \log Z(\tilde R,\tilde r).
\end{equation}
Recall that the partition function has the form
\begin{equation}
    \log Z = S_\nu-2\pi R M + 2\pi R \mu_p Q_p + 2\pi R \mu_w Q_w, \quad S_\nu \equiv S + 2\pi i (1-\nu) J.
\end{equation}
By taking derivatives of \eqref{eq:Z_eq_2}, we can express $S_\nu$, $M$, $Q_p$, and $Q_w$ in terms of the seed solution and the transformation parameters. It is convenient to define the parameters
\begin{equation}
    \tanh \alpha = i \frac{u \tilde R}{\tilde r}, \quad \tanh \beta = i v\tilde R \tilde r,
\end{equation}
in which
\begin{equation}
\begin{split}
    R &= \tilde R \cosh \alpha \cosh \beta +  \frac{n}{\tilde R} \sinh \alpha \sinh \beta\\
    r &= \frac{\cosh \beta }{\cosh \alpha  + \frac{n}{\tilde R^2} \sinh \alpha  \tanh(\beta)} \tilde r .
\end{split}
\end{equation}
Defining 
\begin{equation}
    \tilde S'_\nu = e^{2\tilde \phi_D} \tilde S_\nu, \quad \tilde E' = e^{2\tilde \phi_D} \tilde E,
\end{equation}
 we find
\begin{equation}
\begin{split}
    2\pi Q_p &= e^{-2\phi_D} \frac{\tilde S'_\nu}{\tilde R} r \sinh \alpha \cosh \alpha , \quad 
    2\pi Q_w = e^{-2\phi_D} \frac{\tilde S'_\nu}{\tilde R} \frac{1}{r} \sinh \beta \cosh \beta \\
    S_\nu & = e^{-2\phi_D} \tilde S'_\nu\left(\cosh \alpha \cosh \beta -\frac{n}{\tilde R^2}\sinh \alpha \sinh \beta \right),\\
    M &= e^{-2\phi_D} \left( \tilde M' + \frac{\tilde S'_\nu}{2\pi \tilde R} \left(\sinh^2(\alpha)+\sinh^2(\beta)\right)\right).
\end{split}
\end{equation}
These are the variables used in \cite{Chen:2021dsw}.
For us, it would be useful to change variables again to
\begin{equation}
    \alpha = \frac{\upalpha+\upbeta}{2},\quad 
    \beta = \frac{\upalpha-\upbeta}{2}.
\end{equation}
and $Q_{L,R} = Q_n/r\pm Q_w r$. In terms of the new variables, the transformed properties are
\begin{equation}
\begin{split}
    R &= \frac{\tilde R+n/\tilde R}{2}\cosh \upalpha  +\frac{\tilde R-n/\tilde R}{2}\cosh \upbeta ,\\
    r &= \frac{\tilde R (1+\cosh(\upalpha-\upbeta))}{(\tilde R+n/\tilde R)\cosh \upalpha  + (\tilde R-n/\tilde R)\cosh \upbeta } \tilde r,\\
    Q_L &= e^{-2\phi_D} \frac{\tilde S'_\nu}{2\pi \tilde R}\sinh \upalpha \cosh \upbeta , \quad 
    Q_R = e^{-2\phi_D} \frac{\tilde S'_\nu}{2\pi \tilde R} \cosh \upalpha \sinh \upbeta \\
    S_\nu & = e^{-2\phi_D} \tilde S'_\nu \left(
    \frac{1-n/\tilde R^2}{2}\cosh \upalpha  +\frac{1+n/\tilde R^2}{2}\cosh \upbeta 
    \right),\\
    M &= e^{-2\phi_D} \left( \tilde M' + \frac{\tilde S'_\nu}{2\pi \tilde R} \left(\cosh \upalpha \cosh \upbeta -1\right)\right).
\end{split}
\end{equation}

\bibliographystyle{JHEP}
\bibliography{ref}
\end{document}